\documentclass[a4paper]{elsarticle}
\usepackage{graphicx}
\usepackage{amsmath,amsfonts}
\usepackage{hyperref}
\usepackage{caption}
\usepackage{color}

\newcommand{\R}[0]{\mathbb{R}}

\makeatletter
\newcommand*{\textoverline}[1]{$\bar{\hbox{#1}}\m@th$} 
\makeatother

\makeatletter
\def\ps@pprintTitle{%
 \let\@oddhead\@empty
 \let\@evenhead\@empty
 \def\@oddfoot{\footnotesize\itshape
 \ifx\@journal\@empty Preprint
\else\@journal\fi\hfill\today
}
 \let\@evenfoot\@oddfoot}
\makeatother

\begin{document}

\title{Uncovering inequality through multifractality of land prices: 1912 Kyoto, a case study}

\author[casa]{Hadrien Salat\corref{cor1}}
\ead{hadrien.salat.14@ucl.ac.uk}

\author[crdc]{Roberto Murcio}
\ead{r.murcio@ucl.ac.uk}

\author[ru]{Keiji Yano}
\ead{yano@lt.ritsumei.ac.jp}

\author[casa]{Elsa Arcaute}
\ead{e.arcaute@ucl.ac.uk}

\cortext[cor1]{Corresponding author}

\address[casa]{Centre for Advanced Spatial Analysis, University College London,\\ London, UK}
\address[crdc]{Consumer Research Data Centre, University College London, London, UK}
\address[ru]{Department of Geography, Ritsumeikan University, Kyoto, Japan}

\begin{abstract}
Multifractal analysis offers a number of advantages to measure spatial economic segregation and inequality, as it is free of categories and boundaries definition problems and is insensitive to some shape-preserving changes in the variable distribution. We use two datasets describing Kyoto land prices in 1912 and 2012 and derive city models from this data to show that multifractal analysis is suitable to describe the heterogeneity of land prices. We found in particular a sharp decrease in multifractality, characteristic of homogenisation, between older Kyoto and present Kyoto, and similarities both between present Kyoto and present London, and between Kyoto and Manhattan as they were a century ago. In addition, we enlighten the preponderance of spatial distribution over variable distribution in shaping the multifractal spectrum. The results were tested against the classical segregation and inequality indicators, and found to offer an improvement over those.
\end{abstract}


\maketitle

\section{Introduction}\label{intro}

Reardon et al. \cite{RFOM} pointed out the necessity for new spatial economic segregation and inequality measures insensitive to the choice of category thresholds, boundary definitions (MAUP), and shape-preserving changes in the variable distribution. Despite improving the situation, we have identified that the new measures proposed in the same article still face boundary definition problems and are unsatisfactorily insensitive to all changes in the variable distribution, including non shape-preserving ones. Multifractal analysis could offer a good alternative, free of all the aforementioned problems, for sets obeying a number of scaling conditions.

It is well known that scaling often emerges in urban systems, and the fractality of urban structures is well documented \cite{Fh,BL,B}. We argue that a single fractal exponent may not be sufficient to fully characterize urban systems, and that real estate could be better described by multifractals \cite{HCWX,HCWX2}. The purpose of this article is to make use of two highly detailed land-prices datasets for Kyoto spanning over a hundred years \cite{YNIBS,YNITKMSKTIK,YNIK,YNKT}, and of urban simulated models based on this data to prove that multifractals are indeed a relevant description of the spatial heterogeneities of land price measures in this case.

In addition to the factual results, we aim at setting a general framework for multifractal analysis applied to urban systems, with a focus on interpreting the results in terms of urban inequality and segregation. The information obtained from multifractal analysis is compared in detail to the information resulting from classical inequality and segregation measures (such as Gini, Theil, Neighborhood Sorting, Ordinal Information Theory and Ordinal Variation Ratio indices). We show that the information from classical tools is recovered using multifractal analysis, and that some additional information is gained.

The main factual results are a sharp decrease in multifractality for present Kyoto compared to Taish\textoverline{o} era Kyoto (a trend for modern cities in line with other studies \cite{MMAB,AVHG}), and a striking similarity both between present Kyoto and present London, and between Kyoto and Manhattan as they were a century ago. From the way the multifractal spectrum shrank, we can infer an increase in local homogeneity for the modern city and hints at densification. Furthermore, we show that for this data the resulting spectra are primarily dictated by the shape of the spatial distribution rather that the shape of the price distribution, a characteristic highly sought after by Reardon et al.

We first introduce the methodology and its interpretation for the analysis of inequality, together with a recap of the technical choices made to suit the methods to the current data. We then present the main results, first for Taish\textoverline{o} era Kyoto and its associated simulated models, second for present Kyoto and its associated models, and third we compare the two timestamps between themselves and with data for present Manhattan and London. The third section is a discussion mainly focused on the comparison between the multifractal methodology and classical inequality tools.

\section{Methodology}\label{method}

We start by presenting the heuristics behind multifractal theory and illustrate visually the meaning of the main variables. We then explain how the methodology was adapted to suit a real estate context, and how the data was prepared to mitigate its imperfections. The technical details on the actual computation can be found in \ref{app:multtheo}.

\subsection{Multifractal measures}\label{Mineq}

Multifractal theory \cite{FP,EM,F} can be used to study the heterogeneity and irregularity of measures defined on sets that are too irregular for classical geometric tools, when those measures present two separate scaling properties. Consider a measure $\mu$ defined on a set $A$, it is required that
\begin{enumerate}
	\item locally around any point $x$ of the set $A$, the measure is scaling with a local exponent $\alpha_x$;
	\item the set formed by all points around which the measure scales with the same local exponent $\alpha_x$ is a fractal set of dimension $f(\alpha_x)$.
\end{enumerate}

There are several more or less equivalent definitions of ``fractal'' sets. Here, we will consider \emph{fractal} any set for which a box-counting dimension can be computed (regardless of whether said dimension is a fraction or not, or whether the set is self-similar in a strict sense or not). We will evaluate the multifractality of land prices using the classical moment method \cite{SMA,HJKPS,ASW}, and in particular its multiplier variant (see \cite{CS,C}).

\begin{figure}
      \centering
      \includegraphics[width=\columnwidth,keepaspectratio]{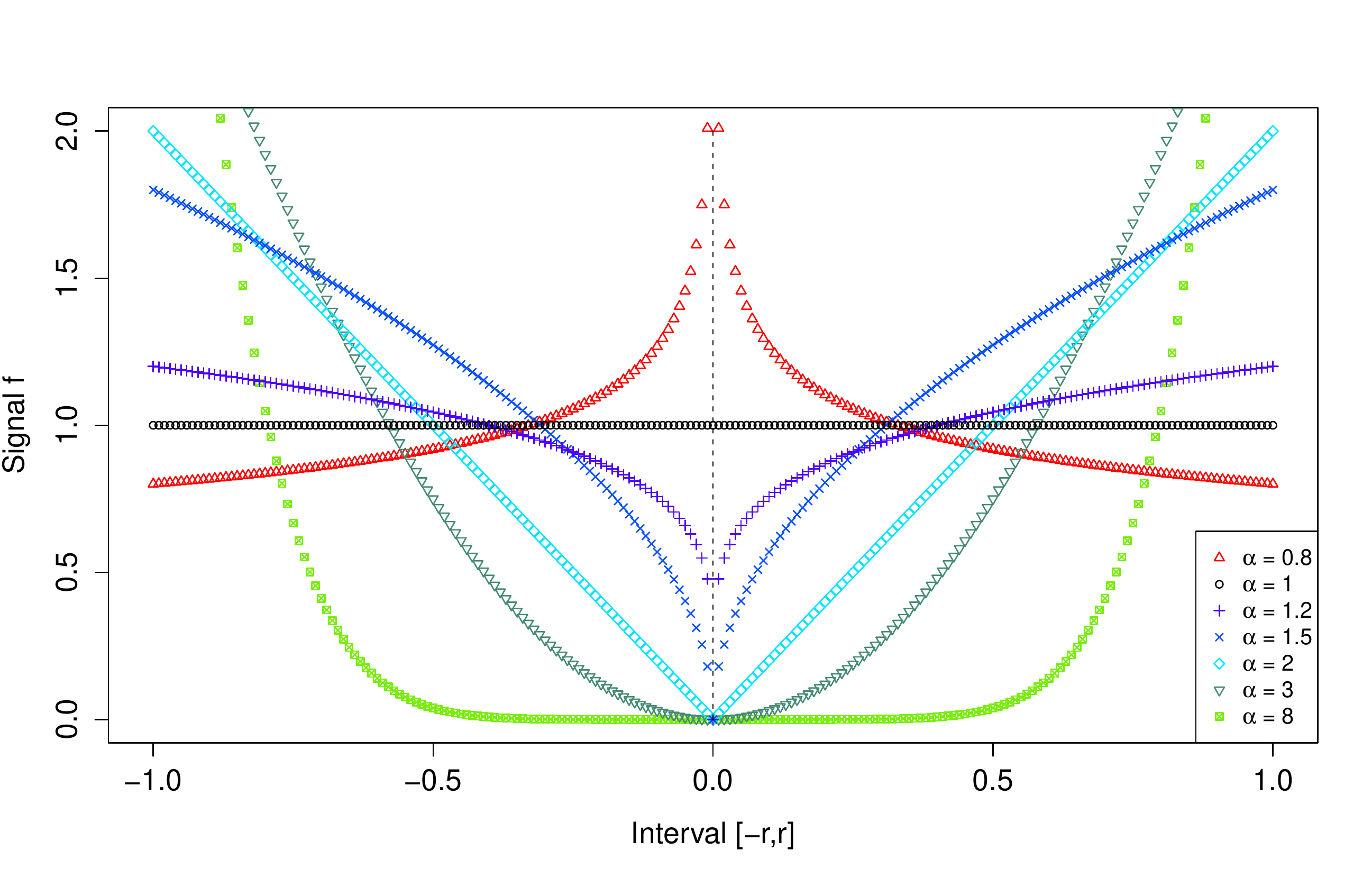}
			\caption{Local behavior of a one-dimensional signal around a point of strength $\alpha$.
			\label{alphaloc}}
\end{figure}

The curve $f(\alpha)$ against $\alpha$ is called the \emph{multifractal spectrum}. It gives, roughly speaking, the ``fractal dimension'' $f(\alpha)$ of sets where the measure scales locally with the same exponent $\alpha$. In order to get a heuristic idea of what the $\alpha$ values represent, assume that we want to study the multifractality of a non-negative signal $\phi$ defined on a one-dimensional space, say $\R$. We can create a measure $\mu$ from this signal by defining for any $a<b\in\R$
\begin{equation}
\mu([a,b))=\int_a^b\phi(x)dx.
\end{equation}
In particular, the first scaling rule $\mu([0,r))\propto r^\alpha$ translates into $\phi(x)\propto x^{\alpha-1}$ on the interval $[0,r)$. Assuming there is a reflectional symmetry around 0, this formalization allows us to illustrate the typical limiting behaviour of a multifractal signal around a point corresponding to some particular $\alpha$ when $r\rightarrow0$ (see figure~\ref{alphaloc}). Note that choosing to place the point in $0$ and the vertical scaling are both arbitrary in this example.

\begin{figure}
      \centering
      \includegraphics[clip,width=\columnwidth,keepaspectratio]{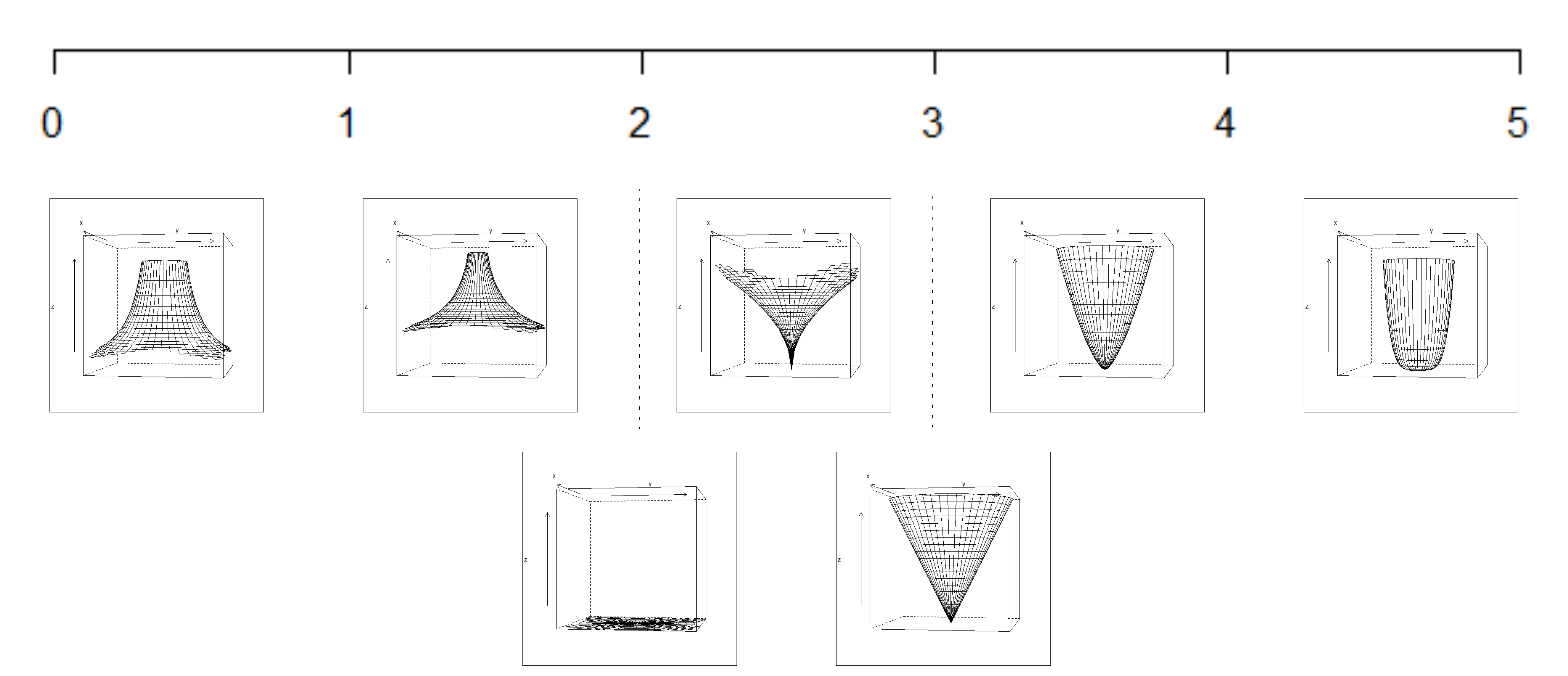}
			\caption{Idealized local behaviour of a two-dimensional signal around a point whose strength $\alpha$ is indicated by the scale on the top.
			\label{alphaloc2D}}
\end{figure}

If we were to consider instead a perfectly isotropic two-dimensional signal, then the cross-cut along any particular direction would be the same as the curves above at the condition of adding 1 to the corresponding $\alpha$. A representation of such a signal for particular $\alpha$ values can be found in figure~\ref{alphaloc2D}. The real signal does not have to be isotropic, for example ``bumps'' on one side of a circle of radius $r_0$ can be compensated by ``holes'' on the other side of the same circle.

It must be emphasized that $\alpha$ values represent the rate at which a signal, and more generally any measure considered, grows around a point in contrast to the actual height of its point value. However, the $\alpha$ values are not completely invariant through vertical translation of the signal. Assume that a signal is of the form $r^{\beta}+k$ for some $\beta$ and a real constant $k>0$. Then, once integrated, the measure will be either of the form $r^{\alpha}+k*r$ or $r^{\alpha}+k*r^2$ for one-dimensional or two-dimensional signals respectively. The noise introduced by $k$ is therefore negligible for $\alpha$ values below $1$ or $2$ (resp.) when $r\rightarrow0$, while the measure is negligible compared to the noise for $\alpha$ values greater than $1$ or $2$ (resp.) when $r\rightarrow0$. As a matter of fact, the standard moment method applied to real data sometimes seems to fail for $\alpha$ values greater than $1$ or $2$ \cite{MMAB,C,HCWX}. In contrast, the multiplier method assumes by construction that the signal would actually reach $0$ (i.e. that the noise $k$ does not exist even when $\alpha>1,2$), and computes an approximation of the $\alpha$ values from the growth of the signal away from the point despite the noise. This process could be thought of as ``calculating the best multifractal fit'' for the measure. The relevance of the spectrum resulting from the multiplier method for $\alpha$ above $1$ or $2$ would depend on how close the data is to a true multifractal.

\begin{figure}
      \centering
      \includegraphics[clip,width=\columnwidth,keepaspectratio]{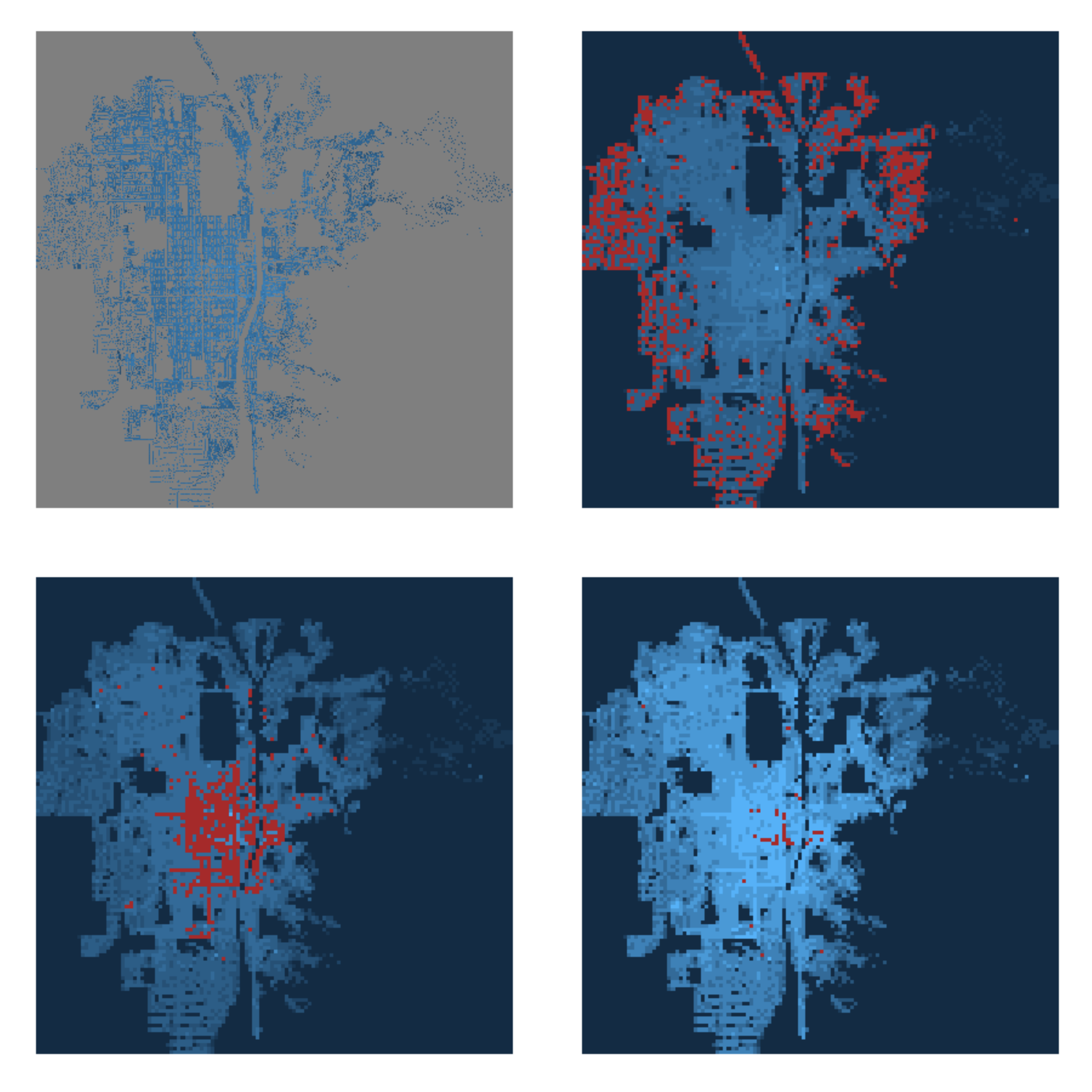}
			\caption{Definition of $f(\alpha)$. Top left represents the price distribution of Kyoto in 1912 in a log-scale, the other figures represent the corresponding $\alpha$ distribution with: top right: $\alpha=3$ highlighted in red, bottom left: $\alpha=6$ highlighted in red, and bottom right: $\alpha\geq7$ highlighted in red. For each $\alpha$ value, the dimension $f(\alpha)$ is akin to the box-counting dimension of the highlighted zone.
			\label{alphagrid}}
\end{figure}

The $f(\alpha)$ values may mean slightly different things depending on how they have been defined. Here, they are assimilated to the box-counting dimension of the set formed by all squares in a grid whose measure shares the same strength $\alpha$. This is illustrated in figure~\ref{alphagrid}. On the top left, the land price distribution for Kyoto in 1912 is represented in a blue continuous logarithmic scale applied over a 512x512 grid. In the three other images, the 4x4 squares corresponding respectively to a strength of $\alpha=3$, $\alpha=6$, and $\alpha\geq7$ (counting clockwise) are highlighted in red. The values $f(\alpha=3)$, $f(\alpha=6)$, and $f(\alpha\geq7)$ are the box-counting dimension of each highlighted zone. The final aim would be to infer from the data the $\alpha$ and $f(\alpha)$ values that would correspond to neighbourhoods of size 0, not 4x4. As a result, the $\alpha$ values in this illustration are largely overestimated compared to the final results in part \ref{result}.

Since lower and higher $\alpha$ values mean sharper growth, while values around $1$ for one-dimensional measures and around $2$ for two-dimensional measures are a sign of local homogeneity (with only isolated singularities), some non-intuitive insights on segregation can be deduced from the repartition of the $\alpha$ values. Indeed, if the $\alpha$ distribution is narrow and centred around $1$ or $2$, while the spread of the measure is high, it means that the prices are clustered into groups of locally similar $\alpha$ values (for which the $f(\alpha)$ dimension is close to 2), while the lower and higher $\alpha$ values represent the sharper edges of the clusters (for which the $f(\alpha)$ dimension is closer to 1). In contrast, if both the measure spread and the $\alpha$ distribution are narrow, then segregation is low.

In particular, it has been evidenced in \cite{MMAB} that there is a loss of multifractality over time in modern cities, which translates into a shrinking of the spread of $\alpha$ values. The remark above encourages to pay extra attention to the symmetries of the shrinking, rather than to its extent alone.

\subsection{Application to measuring land ownership inequality and datasets}\label{Mdata}

The input for the multifractal moment methods needs to be a mathematical measure. As a result, continuous densities cannot be used directly, and a suitable base unit must be chosen to count the measure. For real estate, several units inferring different interpretations can be considered.

To describe the spatial distribution of land value, the unit could be each square meter of land. To describe property value instead of land value, the unit should be based on floor space (one square meter times the number of stories for example). Indeed, considering only house price per square meter would miscalculate the real housing value carried by the land and would underestimate the availability of each particular housing price.

To study housing inequality, each accommodation can be considered as a unit of its own, independently of its size. The results would then give information on the affordability of suitable housing assets and on the spatial repartition of each price category. Adopting the point of view of policy makers, one could choose a set number of people as the base unit and try to calculate the cost of accommodating that number of people.

Unfortunately, the data available may have the final word on deciding the unit. For early 20th century Kyoto, we have based our work on the digitalization of a cadastral map first published in 1912 (first year of Taish\textoverline{o} era) by Yano et al. \cite{YNIBS,YNITKMSKTIK,YNIK,YNKT}. At the smallest level, all land is divided into lots registered to a single owner for which price, category, and area are known. We used the total price of each lot marked as residential as the base unit for the analysis. Assuming that the land lots are not meant to be divided, it represents accurately all existing residential land assets in the city. We could have considered lots of categories other than residential to represent assets for future development of the city, but we thought it would be more consistent to focus exclusively on built land. For the record, non-residential land represents $18\%$ of all unique lots.

For present Kyoto, the data available is presented as land tax assessed value by road valuation around 2012, provided by Kyoto City. The price is given as mean land price per square meter along each road segment, where a road segment is defined as any part of a road included between two street intersections (or dead ends). Most road segments are edges of a block, and some are smaller intra-block streets. Making the unavoidable assumption that the difference in depth between different residential buildings is negligible compared to the length of the roads segment, we have multiplied the mean price along each segment by the length of said segment. The result are values proportionate to the sum of all land lot prices along each segment. Since multifractals are invariant under a linear transformation, those values can be seen as approximations comparable to 1912 lot values (although at a broader scale). Both the full extent of modern Kyoto and its intersection with the extent of 1912 Kyoto will be considered. All study areas are represented in figure~\ref{studyareas} of \ref{app:CityModel}.

For both datasets, values have been paired with the centroids of their corresponding lots or road segments. The data is processed as a matrix representing a grid overlay over the Kyoto map. Due to the relatively broad scale (especially for present Kyoto), to the interference created by simplification during the price assessment process, and to the fuzzy nature of real estate, we do not expect that the data will be a clean multifractal with definite self-similarity at all scales. For that reason we use the multiplier method to calculate in a sense the ``best fit'' spectrum. Even if the data is not exactly self-similar at all scales, the multifractal methodology allows nonetheless a consistent comparison of the variety and spatial partitioning of prices between different city models as long as the data shares a similar resolution and structure. The results for present Kyoto are to be taken with more caution than those for Taish\textoverline{o} era Kyoto because of the difference in data quality. Similarly, as evidenced in the previous section, the part of the spectra for $\alpha$ values below 2 is more reliable and expected to be in better accordance with more ``rigid'' multifractal methods than the part for $\alpha$ values greater than 2.

\section{Results}\label{result}

Urban models were used to analyse the influence of changing the spatial distribution or the price distribution on the resulting multifractal spectra. We first present a detailed analysis for Taish\textoverline{o} era Kyoto and its associated models, then a more concise analysis for present Kyoto due to the less appealing shape of the data. We finally make a temporal comparison between both datasets, and use the PLUTO database for New-York and Land Registry house transactions database for London to compare the results with these two cities. Technical aspects of the urban models creation and isolated spectra for each type of model are provided in more detail in \ref{app:CityModel}. While we focus here on the multifractal spectra alone, the so-called generalized dimension usually associated to multifractal analysis is indicated in \ref{app:D0D1D2}.

\subsection{City models: 1912 data}\label{RCM1912}

\begin{figure}
      \centering
      \includegraphics[clip,width=\columnwidth,keepaspectratio]{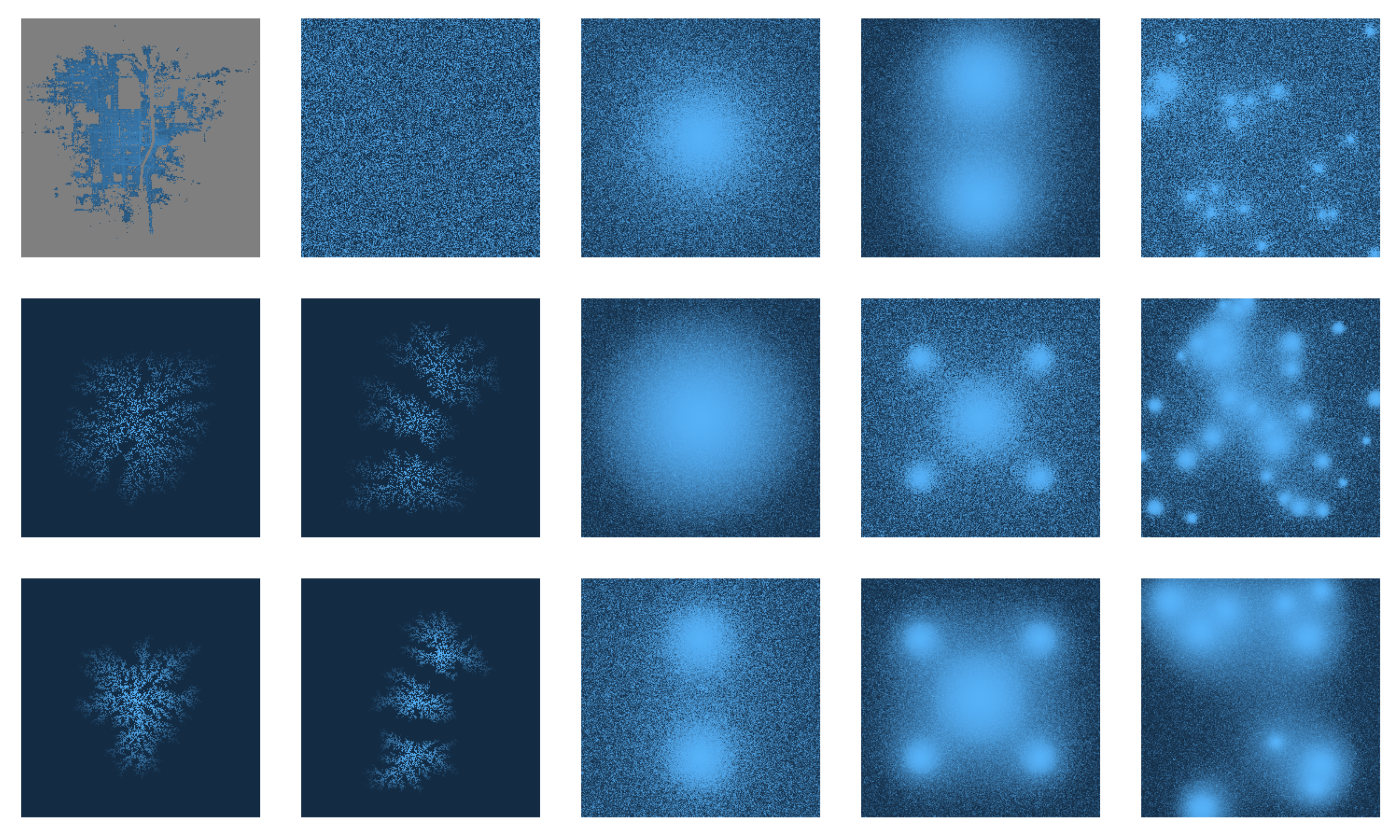}
			\caption{Top left: logarithmic overlay of the real price distribution over the real Kyoto pattern in a grid of resolution 256x256. Next to it: distribution ranks drawn uniformly. Four figures on the bottom left: rank overlay over DLA models with 1 or 3 centres and different levels of attraction. Nine figures on the right: rank overlay over polycentric models with different number of centres and levels of attraction.
			\label{abstmodspace}}
\end{figure}

A first batch of models is built by changing the spatial pattern while the price distribution is kept identical to the true distribution. Three types of spatial distributions are chosen to supplement the true pattern: uniform, polycentric, and diffusion-limited aggregation (DLA). The first one is used as a null model, the second one as a representation of how modern megacities develop \cite{O2,MS,LB}, and the third one as a multifractal reference since it is known to generate strong multifractality \cite{WS,VFM,MRR,RRM}. All these models are plotted in figure~\ref{abstmodspace} alongside the true distribution. The latter is represented in the top left image. Next to it is the price distribution drawn uniformly. The four figures on the bottom left are DLA models with either 1 or 3 centres exerting different levels of attraction. Finally, the nine models on the right are polycentric models with different number of centers exerting varying levels of attraction. The true spatial distribution uses a logarithmic overlay of the real price distribution, while all other images use an overlay representing the rank of each point in the price distribution after it has been drawn into the space. All images are represented in a grid of resolution 256x256.

A second batch of models is built by changing the price distribution while the spatial pattern is kept identical. Three distributions are considered: uniform, truncated normal, and Pareto. The uniform distribution is taken as a null model, while the normal and Pareto distributions are the most recurrent distributions observed in urban science (in particular, the Pareto distribution is usually associated to the distribution of wealth). The true distribution is log-normal. To create the distributions, range, mean and standard deviation can be adjusted. Since it is by construction impossible to match all three parameters for each distribution, priority was given to matching the range with the true distribution. When the exceptionally high priced imperial palace is removed, this range consists of values between 0 and 10000 yens. These three distributions are plotted in \ref{app:CityModel}. In addition to the price distributions being laid over the true spatial pattern, they have also been laid over the uniform spatial distribution for reference.

\begin{figure}
      \centering
      \includegraphics[clip,width=\columnwidth,keepaspectratio]{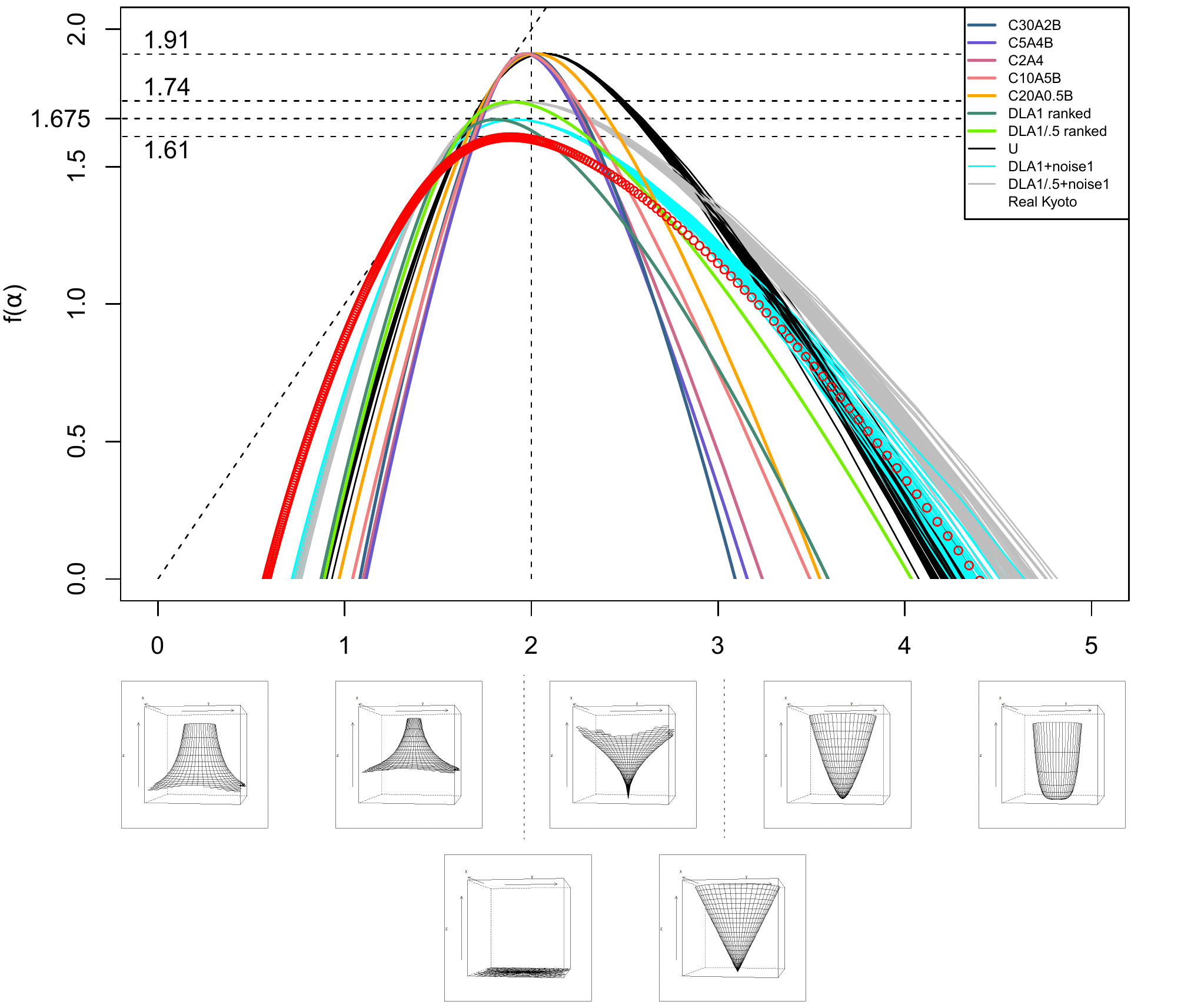}
			\caption{Transversal representation of multifractal spectra for Kyoto 1912 price distribution mapped over several spatial models. Corresponding idealized two-dimensional signal below the $\alpha$ axis was added for reference.
			\label{transvers}}
\end{figure}

The spectra resulting from the first batch of models plotted against the true Kyoto distribution can be seen in figure~\ref{transvers}. Only the most relevant cases were selected, and some additional curves for each type of model can be found in \ref{app:CityModel}. We have in total: six polycentric models (referred to as $CxAy(B)$ in the legend, where $x$ is the number of centres, $y$ a coefficient of attraction, and $B$ indicates that the centres are weighed differently), fifty iterations of the uniform draw ($U$ in the legend), and two fully ranked DLA models with different centre attraction weights and fifty iterations of each DLA model with a noise added to it (respectively $DLA1 ranked$, $DLA1/.5 ranked$, $DLA1+noise1$, and $DLA1/.5+noise1$). We can immediately observe that the real distribution (red dots) seems to maximize the width of the spectrum, while minimizing its height. Only some of the DLA models with added noise could create a wider right-spectrum and remain close on the left side. Polycentric models generate fairly weak multifractality, especially considering that the multiplier method tends to artificially widen the spectrum in weak cases. The uniform distribution lays between the polycentric and DLA distributions.

\begin{figure}
      \centering
      \includegraphics[clip,width=\columnwidth,keepaspectratio]{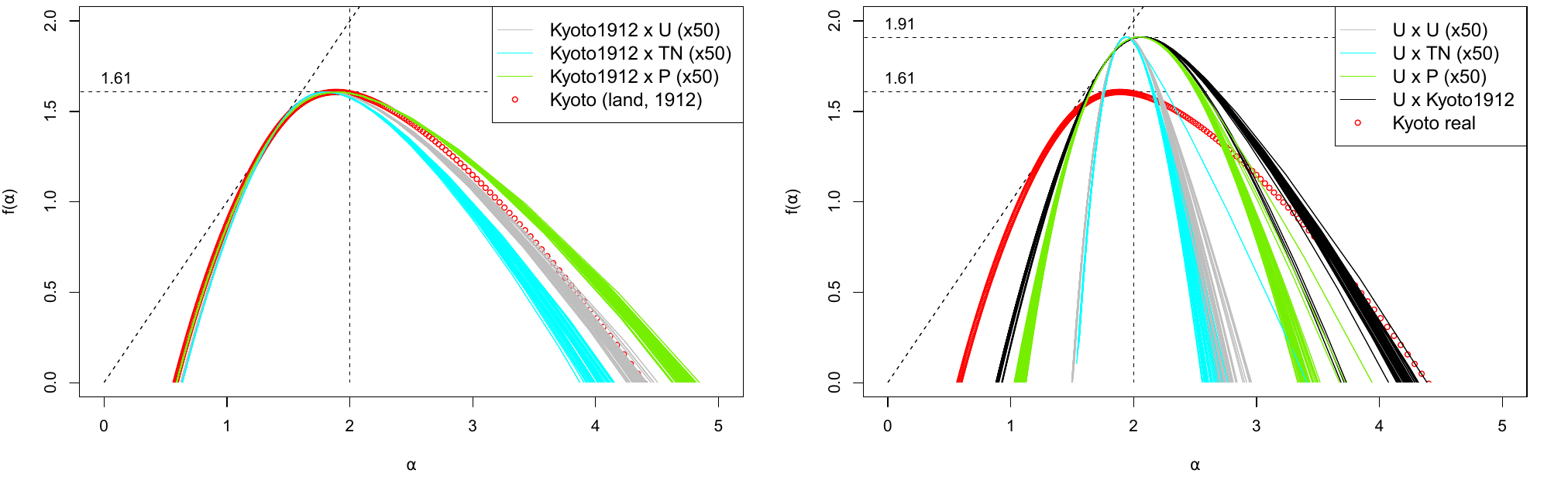}
			\caption{Uniform (grey), truncated normal (teal), Pareto (green) and real (red) distributions mapped over 1912 Kyoto true spatial distribution on the left, and uniformly drawn into space on the right.
			\label{KUxAll}}
\end{figure}

Since the price distribution is the same for all models, we can conclude that the spatial distribution has a large impact on shaping the width of the spectrum. This is further corroborated by the second batch of models for which several price distributions are mapped over the true distribution and over the uniform distribution (see figure~\ref{KUxAll}). In the first case, there is almost no difference in the left part of the spectra, while the right part are very similar, with Pareto distribution giving the widest spectrum. However, the price distribution does have an impact on the spectrum, as evidenced by the second case, where all four price distributions have been drawn randomly in space. The uniform and truncated normal distributions present almost no multifractality, the very narrow spectrum being identifiable to an artefact of the multiplier method. The Pareto and the real distributions produce more convincing spectra, close to one another.

\subsection{City models: 2012 data}\label{RCM2012}

\begin{figure}
      \centering
      \includegraphics[clip,width=\columnwidth,keepaspectratio]{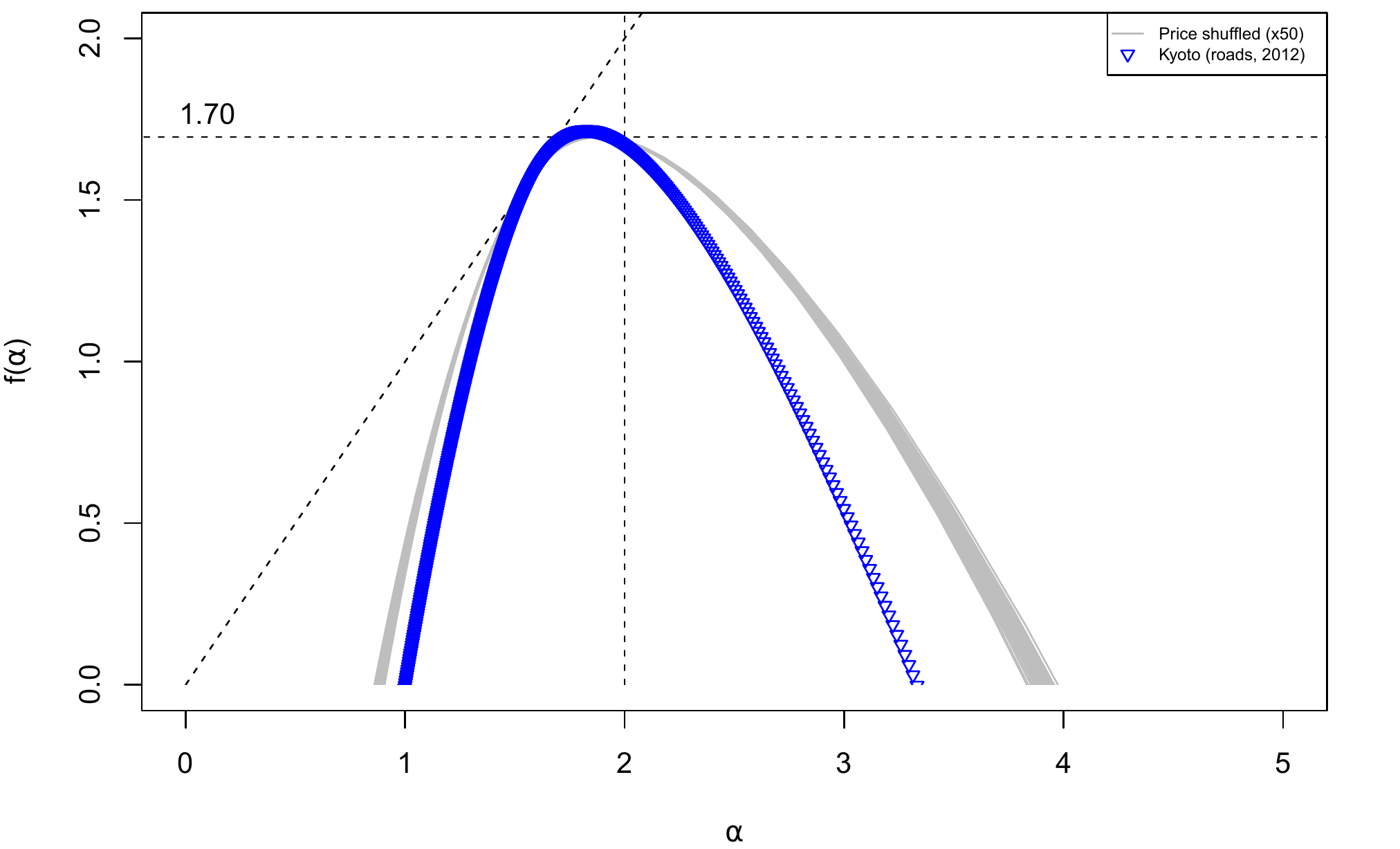}
			\caption{Kyoto 2012 price distribution shuffled inside the true spatial pattern (grey) compared to the true 2012 distribution (dark blue).
			\label{K2012shuf}}
\end{figure}

Recall that for present Kyoto, the processed data consists of centroid coordinates for each road segment in Kyoto. To each centroid is attributed the mean square meter price multiplied by the length of each corresponding segment. Instead of constructing many less reliable and redundant models compared to the ones built for Taish\textoverline{o} era Kyoto, we focus on shuffling randomly the price distribution in a space where all the actual locations of road segment centroids are preserved. Fifty iterations of this process were done and produced the spectra that can be seen in figure~\ref{K2012shuf}. Contrary to the 1912 case, the real placement seems to minimize the spectrum width.

\begin{figure}
      \centering
      \includegraphics[clip,width=\columnwidth,keepaspectratio]{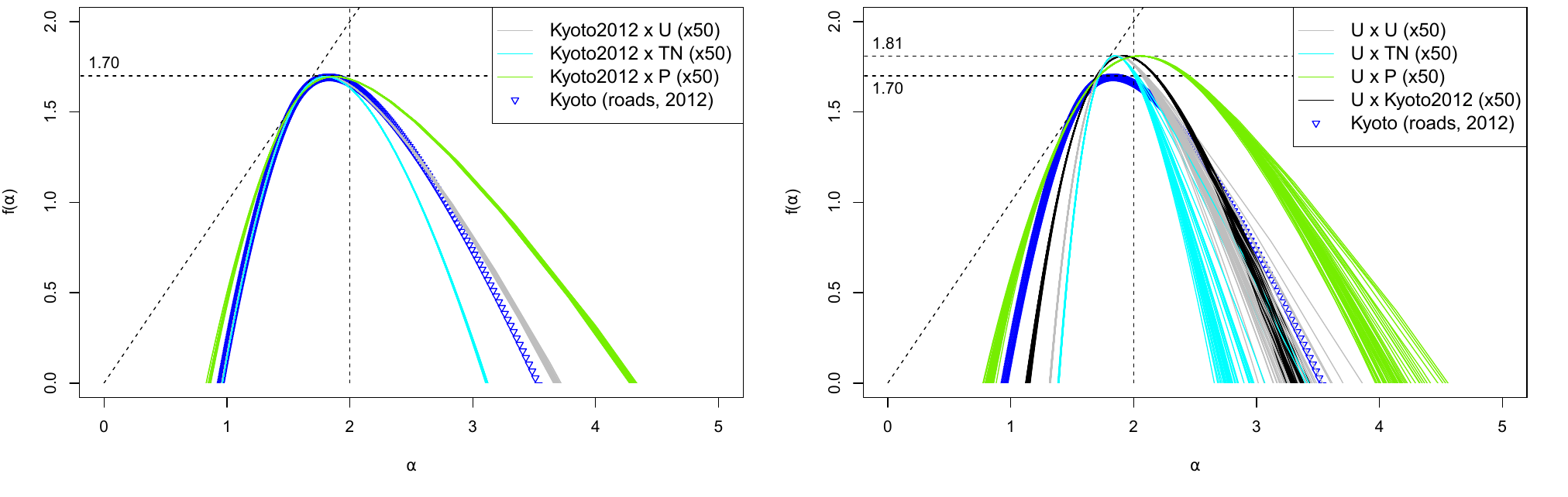}
			\caption{Uniform (grey), truncated normal (teal), Pareto (green) and real (red) distributions mapped over Kyoto true spatial distribution in 2012 on the left, and uniformly drawn into space on the right.
			\label{KUxAll2012}}
\end{figure}

The price distribution for 2012 Kyoto is almost log-normal, akin to the price distribution for 1912, although it is less symmetrical and only spans over two orders of magnitude compared to four in the previous case. Similarly, we generated 50 iterations of some corresponding uniform, normal and Pareto distributions (see \ref{app:CityModel} for a graphical representation). The spectra for each distribution can be found mapped over the true spatial pattern in the left panel of figure~\ref{KUxAll2012}, and uniformly drawn into space in the right figure of figure~\ref{KUxAll2012}. The results are similar to those obtained for 1912. In the true pattern case on the left, the spectra follow the same ranking in width as for 1912, with only slightly more pronounced differences. In the uniform pattern case on the right, the actual price distribution gives results closer to the uniform distribution, while the Pareto distribution provides significantly wider results. This is coherent with the narrower range and higher maximum of the true price distribution, reflected in a more dispersed corresponding Pareto distribution.

\subsection{Comparing Kyoto across time, and with data from other cities}\label{RK}

To be able to compare accurately Kyoto in 1912 and in 2012, it would have been preferable to use the same resolution in both cases. Unfortunately, the road valuation data for 2012 creates undesired ``gaps'' between data points when plotted at the resolution used in the 1912 case. Those gaps do not describe real land value well, as price lots should be contiguous or almost contiguous, with only big natural obstacles (such as the Kamo river for Kyoto) representing an actual empty space. To tackle this problem, the resolution used for present Kyoto has been reduced by 4 to obtain a relatively compact grid. A first map was created representing the full extent of the present city at a resolution of 512x512, the same resolution that was used for the 1912 extent of the city, despite the new city being roughly four times bigger than the old city. A second map was created to represent the evolution of the old city alone. It comprises only the roads that are fully included in the boundaries of the 1912 city. This map is at a resolution of 216x216, i.e. four times lower than for the 1912 map, despite the geographical extent being identical. The study areas for each case are illustrated in figure~\ref{studyareas} of \ref{app:CityModel}.

\begin{figure}
      \centering
      \includegraphics[clip,width=\columnwidth,keepaspectratio]{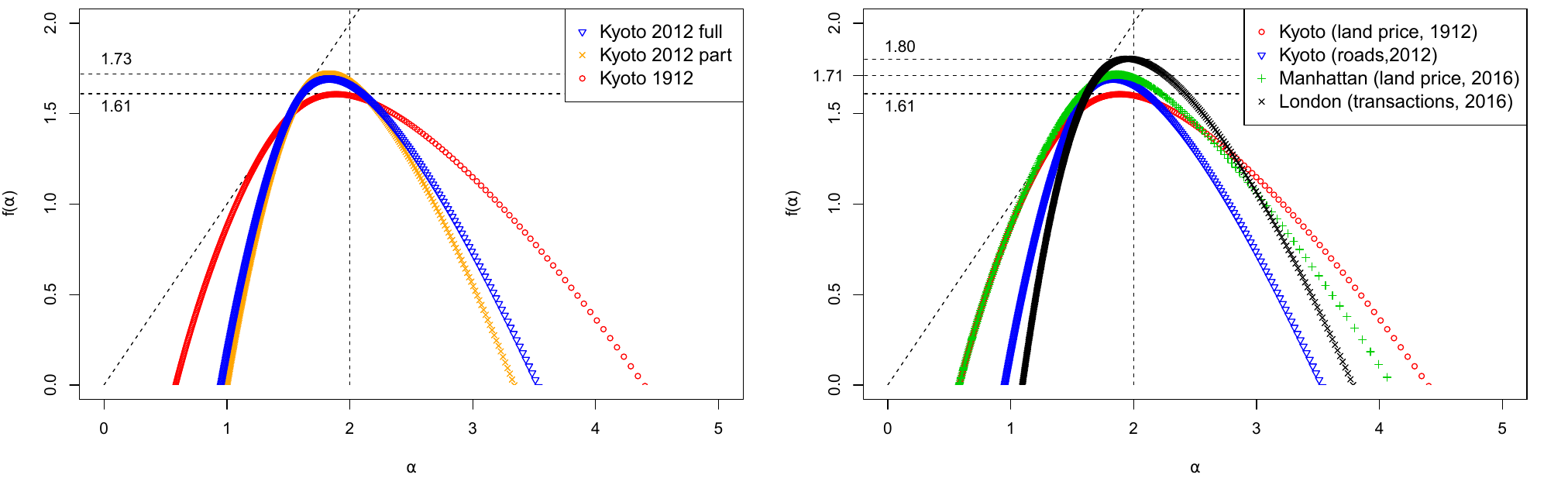}
			\caption{Left: Comparison between Kyoto price distribution in 1912 and in 2012. Kyoto 2012 part indicates that the extent of modern Kyoto was limited to its 1912 boundaries. Right: Comparison between Kyoto land price distribution in 1912, Kyoto road valuation in 2012, Manhattan land price distribution in 2016, and London house transaction prices in 2016.
			\label{Comp_london_mn}}
\end{figure}

The spectra for both maps are plotted against the 1912 spectrum in the left panel of figure~\ref{Comp_london_mn}. It could be hypothesized that the increase in spectrum height for the modern city is representative of densification. Indeed, the eastern half of the city is much more compact in 2012 than it was in 1912. Unfortunately, it can also be an artefact of the lower resolution, so no definite conclusions can be drawn.

The evolution of the width of the spectrum is more interesting. Since summing measures defined on disjoint supports will result in a spectrum for which $f(\alpha)$ is the maximum of the $f(\alpha)$ in the spectrum of each independent measure, it was expected that the spectrum corresponding to the 1912 boundaries map would be narrower than the spectrum for the full present city. Adding newly developed zones to the old centre can only add variety and make the spectrum wider. The fact that it is only slightly narrower indicates that there is no significant discrepancy in the development of new neighbourhoods compared to how the old city evolved.

Comparing the 2012 spectrum to the 1912 spectrum, there is an observable loss of extreme $\alpha$ values, corresponding to the steepest spatial increase in price, while the relatively homogeneous zones, i.e. those around $\alpha=2$, are of higher $f(\alpha)$ dimensions in the 2012 case. This is representative of a noticeable increase in local homogeneity. It can also be noted that the shape is more symmetric in 1912 compared to 2012, with a rounder left half of the spectrum. It means that the diversity in 2012 is created more by (small) local ``bumps'', and less by (small) local ``gaps'' compared to 1912.

The 2016 PLUTO database for New-York contains the assessed tax value for each lot in Manhattan for the 2017 fiscal year. According to the documentation, it is ``calculate[d] by multiplying the tax lot’s estimated full market land value, determined as if vacant and unimproved, by a uniform percentage for the property's tax class'' by the Department of Finance. It is quite similar to the assessed land lot value from the Kyoto 1912 dataset. The resulting spectrum can be found in the right panel of figure~\ref{Comp_london_mn}. It can be observed that it is surprising close in width to the spectrum for 1912 Kyoto, particularly on the right side. It is also similar in height to 2012 Kyoto. A look at the values from year to year in the PLUTO database (which starts in 2002), indicates that the prices are quite stable through time, with only small readjustments from one year to another. It can be hypothesize that the 2017 assessed land prices distribution is in reality close to the early 20th century's distribution, explaining the similarity with 20th century Kyoto. On the other hand, the $D_0$ value representing the fractal dimension of the physical city, is closer to present Kyoto.

Another dataset containing all the house transactions that happened in London in 2016 can be obtained from the Land Registry. Although house transactions are not directly comparable to land prices, it gives an idea of the real estate market in London for that particular year. The resulting spectrum is plotted in figure~\ref{Comp_london_mn}. While Manhattan data produced a spectrum close to Taish\textoverline{o} era Kyoto, the data for London in 2016 produces a result that is convincingly close to present Kyoto, particularly for the left (most stable) half of the spectrum. This trend of multifractal loss in modern cities has already been observed for road networks \cite{MMAB,AVHG}.

Overall, the wide spectrum obtained for the real cases and the ability to discriminate between situations, indicate that the multifractal methodology can provide informative insights when applied to land prices measures. It is in accordance with Hu et al. \cite{HCWX} who also found multifractality for land price distributions in Wuhan City in China. We will show in the next section that the results agree with classical inequality indicators, legitimating its use in an urban inequality context.

\section{Discussion}\label{discu}

We compare the results obtained from multifractal analysis with the results obtained when the classical inequality and segregation indicators are applied to the data. We also point out the necessity to contextualize the analysis with inter-measure comparison instead of inter-city comparison alone.

\subsection{Comparison with classical indicators}\label{Dcomp}

In their review of existing economic segregation measures \cite{RFOM}, Reardon et al. distinguish three different groups of measures. The first group, category-based measures, does not take into account the ordinal nature of the variable, and is irrelevant to our study (see \cite{AMPFA} for an extensive compilation of such indices). The second group, variation-ratio measures, such as the Neighbourhood Sorting Index (NSI) \cite{J}, compares the ratio of the between-neighbourhood variation compared to the total population variation for some definitions of variation. The third group, spatial measures, such as an adapted version of geographical autocorrelation \cite{C2}, takes into account the spatial patterning of the variables. It is the most interesting and least well developed one according to them.

The multifractal methodology would pertain to this last group and would not be concerned with the flaws underlined by Reardon et al for the other measures. Indeed, it is not subject to the arbitrary nature of the definition of geographical unit or categorical thresholds, and it is insensitive to linear transformations (and moderately sensitive to affine transformations using the multiplier method) in the variable distribution. To avoid these flaws, Reardon et al. had developed new spatial measures based on the variation-ratio idea, among which are the Ordinal Information Theory Index (OITI) and Ordinal Variation Ratio Index (OVRI). We will try to demonstrate the advantages of the multifractal approach over these measures.

In addition to the segregation NSI, OITI and OVRI indices, we consider three a-spatial inequality measures: the relative dispersion (RD), the popular Gini coefficient, and the Theil index \cite{S}. The formal definitions for all six indices can be found in \ref{app:CI}. For the practical computation, Kyoto is arbitrarily divided in 100 square neighbourhoods.

The results for 1912 Kyoto are shown in table \ref{tab:CI1912P} for the full residential land price distribution (Tot.), for a partial price distribution (Part.) that excludes a few outstandingly expensive lots (such as the Imperial palace), as well as for the uniform (Unif.), truncated normal (TN) and Pareto (Pareto) price distributions mapped over the true spatial distribution. In accordance with the multifractal analysis, the RD, Gini, Theil and NSI coefficients provide similar values between the partial and Pareto distributions on the one side, and between the truncated normal and uniform distribution on the other side. Also in accordance with the multifractal analysis, the RD and Gini coefficients indicate more distributional variety for the partial and Pareto distributions than for the uniform and truncated normal ones. According to a remark in \ref{Mineq}, this translates into higher levels of spatial segregation for the truncated normal and uniform distributions, which is indicated both by the spectra of figure~\ref{KUxAll} and the values of the NSI coefficient. It is noteworthy that the NSI coefficient is made irrelevant for the full price distribution. Indeed, its definition offers no counter to the Imperial palace making all other lots negligible compared to it. In contrast, multifractal analysis is more resilient to unique outstanding value, which can only add to the total variety. Finally, the OITI and OVRI indicate no difference between all price models. This insensitivity is by construction. We prefer the multifractal analysis ability to pick up differences between price distributions even if the spatial pattern is the dominant element determining the shape of the spectrum.

\bigskip

\noindent\begin{minipage}{\textwidth}
	\makebox[\textwidth][c]{\small
		\begin{tabular}{@{}|r|c|c|c|c|c|}
		\hline
			& Tot. & Part. & Price: Unif. & Price: TN & Price: Pareto \\ \hline
			RD & 0.998 & 0.904 & 0.498 & 0.402 & 0.905 \\ \hline
			Gini & 0.666 & 0.606 & 0.332 & 0.281 & 0.610 \\ \hline
			Theil & 1.833 & 0.786 & 0.192 & 0.133 & 0.734 \\ \hline
			NSI & 0.033 & 0.234 & 0.500 & 0.502 & 0.338 \\ \hline
			OITI & 0.147 & 0.147 & 0.147 & 0.147 & 0.147 \\ \hline
			OVRI & 0.166 & 0.165 & 0.166 & 0.166 & 0.166 \\ \hline
		\end{tabular}}
	\captionof{table}{Classical inequality measures for different price distributions (Kyoto, 1912).}
	\label{tab:CI1912P}
\end{minipage}

\bigskip

Another set of results is presented in table \ref{tab:CI1912S}. The partial 1912 distribution has been mapped over the uniform spatial distribution (Unif.), over one center and five centres polycentric models (C1A4 and C5A1B3), and over one seed and three seeds DLA models (DLA1 and DLA3). None of the indicators show segregation for the uniform distribution, which is expected. OITI and OVRI failed to distinguish between C1A4, DLA1 and DLA3 models, and NSI between C1A4 and DLA1 models. The C5A1B3 model shows significantly less segregation than the C1A4 model because it is made roughly of 5 copies of the same concentric distribution, even though the repartition would feel identical from an inhabitant perspective. This shows that these measures are not completely free from the modifiable areal unit problem (MAUP), contrary to the multifractal analysis.

\bigskip

\noindent\begin{minipage}{\textwidth}
	\makebox[\textwidth][c]{\small
		\begin{tabular}{@{}|r|c|c|c|c|c|c|}
		\hline
			& Part. & Space: Unif. & C1A4 & C5A1B3 & DLA1 & DLA3 \\ \hline
			NSI & 0.234 & 0.043 & 0.530 & 0.396 & 0.451 & 0.241 \\ \hline
			OITI & 0.147 & 0.004 & 0.480 & 0.170 & 0.488 & 0.451 \\ \hline
			OVRI & 0.165 & 0.002 & 0.442 & 0.155 & 0.511 & 0.476 \\ \hline
		\end{tabular}}
	\captionof{table}{Classical inequality measures for different space distributions (Kyoto, 1912).}
	\label{tab:CI1912S}
\end{minipage}

\bigskip

The results for 2012 Kyoto are shown in table \ref{tab:CI2012} for the entire dataset (Tot.), for the part of the dataset that corresponds to the extent of the 1912 city (Part.), for a shuffling of the full dataset (Shuffled), and for the uniform (Price: Unif.), truncated normal (Price: TN), and Pareto (Price: Par.) distributions mapped over the real city. As found previously and as expected, the OITI and OVRI coefficients are invariant through change of price distribution, while the RD, Gini and Theil coefficients are invariant through shuffling. Overall, the same decrease of inhomogeneity between 1912 and 2012 is picked up by all coefficients, especially the spatial ones (NSI, OITI, OVRI). Contrary to the multifractal analysis, the shuffled distribution appears more homogeneous than the real distribution, which is true at the broader scale used to compute the NSI, OITI and OVRI coefficients, but not at the microscale used for multifractal analysis. Also contrary to the multifractal analysis, the partial distribution is slightly more unequal than the total one. This is due to the property of multifractal analysis that the spectrum of the total measure must be wider than the spectrum of a part of it. In a sense, multifractality ``adds'' all the variety in the measure whereas the classical indicators are ``rescaled'' inside each subset. Finally, the ranking between price distributions is similar to the one found in figure~\ref{KUxAll2012}.

\bigskip

\noindent\begin{minipage}{\textwidth}
	\makebox[\textwidth][c]{\small
		\begin{tabular}{@{}|r|c|c|c|c|c|c|}
		\hline
			& Tot. & Part. & Shuffled & Price: Unif. & Price: TN & Price: Par. \\ \hline
			RD & 0.701 & 0.768 & 0.701 & 0.498 & 0.282 & 0.991\\ \hline
			Gini & 0.479 & 0.520 & 0.479 & 0.332 & 0.199 & 0.661 \\ \hline
			Theil & 0.400 & 0.479 & 0.400 & 0.192 & 0.067 & 1.02 \\ \hline
			NSI & 0.267 & 0.376 & 0.044 & 0.251 & 0.260 & 0.203 \\ \hline
			OITI & 0.041 & 0.074 & 0.002 & 0.041 & 0.041 & 0.041 \\ \hline
			OVRI & 0.043 & 0.079 & 0.002 & 0.043 & 0.043 & 0.043 \\ \hline
		\end{tabular}}
	\captionof{table}{Classical inequality measures for different space distributions (Kyoto, 2012).}
	\label{tab:CI2012}
\end{minipage}

\bigskip

Overall, almost all conclusions that could be drawn from the classical indicators are in accordance and could also be found with the multifractal analysis. The latter also allowed us to obtain more information and was found to be better at distinguishing some of the spatial models (in particular the polycentric one) and at giving a result more coherent with the intuitive perception of those places.

\subsection{The need to relate results to other measures}\label{Dom}

Contrary to intuition, a narrower spectrum, which means less variety, is not necessarily equivalent to less inequality. As a matter of fact, a narrow spectrum with a wide price range means important segregation, since it indicates that the many price bands are all grouped in locally relatively homogeneous environments. Furthermore, it could appear better suited that the land price distribution matches the income distribution to ensure maximum affordability instead of a situation where all prices are equally unaffordable. Unfortunately, income data tends to be scarcer and usually comes divided into broad categories that are unsuitable for multifractal analysis. Nonetheless, the multifractal results are in accordance with recent research suggesting that segregation in modern Japanese cities is expected to be particularly low. Some social explanations (that are beyond the scope of this article) can be found in \cite{FCH}, an article that was thoroughly reviewed by Fielding in \cite{Fd}.

It would also be interesting to relate accessibility to land prices. Whether a wider or narrower spectrum should be aimed for accessibility would depend on the variable considered: a node network with a wider spectrum offers more option to travel from one point to another with effective hubs, while a time length network should be as narrow and centred over $\alpha=2$ as possible to avoid exponentially long journeys to particular places.

\section{Conclusions}\label{conclu}

We found that the 1912 true distributions tend to maximize the width of the spectrum while the 2012 distributions tend to minimize it. From the shape of the spectra, it appears that the loss of multifractality is representative of more homogeneity in the present city. We found some striking correspondence between early century Kyoto and Manhattan land lots, and between present Kyoto and modern London.

From the comparison with classical inequality and segregation indicators, it appears that the multifractal methodology can offer a valuable complement to the already existing tools, while being free of some of their flaws.

\section*{Acknowledgments}


\paragraph{Funding}{R.M. was supported by ERC Grant [249393-ERC-2009-AdG] and also acknowledges the support of UK ESRC Consumer Data Research Centre (CDRC) [ES/L011840/1], E.A. was supported by EPSRC Digital Economy Phase 2: UK Regions Digital Research Facility (UK RDRF) [EP/M023583/1, 2015–2020].}

\paragraph{Competing interests}{We declare we have no competing interests.}

\paragraph{Acknowledgments}{We acknowledge Dr. Clementine Cottineau for her guidance in part \ref{Dcomp}.}

\paragraph{Data availability}{The Land Registry database for London and the PLUTO database for New-York are open-access and available online as of 2017. A subset containing the coordinates and land prices for Kyoto will be available as supplementary material.}

\appendix

\setcounter{figure}{0}
\setcounter{table}{0}

\section{Technical aspects of the multifractal calculations}\label{app:multtheo}

We recall that a multifractal measure $\mu$ defined on a set $A$ must present two separate scaling properties.
\begin{enumerate}
	\item locally around any point $x$ of the set $A$, the measure is scaling with a local exponent $\alpha_x$;
	\item the set formed by all points around which the measure scales with the same local exponent $\alpha_x$ is a fractal set.
\end{enumerate}

In our framework, that is using the multiplier moment method, denoting $\mu_r(x)$ the measure in a ball of radius $r$ around $x$ and $N(\alpha)$ the number of times an $\alpha_x$ falls inside the interval $[\alpha,\alpha+d\alpha]$, the two scaling properties mentioned above can be written
\begin{enumerate}
	\item $\mu_r(x) \propto r^{\alpha_x}$ for some $\alpha_x$ around any $x\in A$ when $r$ is small enough;
	\item $N(\alpha_x) \propto r^{-f(\alpha_x)}$, for some function $f$ and any $x\in A$.
\end{enumerate}

The goal is to find all the $\alpha_x$ and $f(\alpha_x)$ values in the system. To this end, we apply the moment method. The data is divided into a grid whose squares are numbered. Denoting $\mu_i(r)$ the measure of a Moore neighbourhood of radius $r$ around square $i$, the quantity $Z(q,r)$ is defined for any real number $q$ by
\begin{equation}
\begin{split}
Z(q,r):=\sum_i \mu_i(r)^q &\propto \sum_i r^{\alpha_i q}\\
                      &\propto \sum_\alpha N(\alpha)r^{\alpha q}\\
									    &\propto \sum_\alpha r^{\alpha q-f(\alpha)}.
\end{split}
\end{equation}
For each $q$, when $r\rightarrow 0$, only the value of $\alpha$ that minimizes $\alpha q-f(\alpha)$ makes a significant contribution to the sum, so that the $\alpha$ and $f(\alpha)$ values can be deduced from a Legendre transform of the quantity
\begin{equation}
\tau(q):=\alpha(q) q-f(\alpha(q))\approx\lim_{r\rightarrow0}\frac{\log(Z(q,r))}{\log(r)}.
\end{equation}
Going through all $q$ in $\mathbb{R}$ yields all $\alpha$ and $f(\alpha)$ values. When using the multiplier and gliding box variant, boxes of some minimal size $r_0$ and some other size $r_k$ are glided along the grid. The $\tau(q)$ and $\alpha(q)$ values are then directly computed from the expressions
\begin{gather}\label{taumultp}
\tau(q)+d\approx-\frac{\log\left(1/N\sum_{i}M_{i}^q\right)}{\log(r_k/r_0)};\\
\alpha(q)\approx-\frac{\sum_{i}M_{i}^q\log(M_{i})}{\sum_{i}M_{i}^q\log(r_k/r_0)},
\end{gather}
where $i$ is the grid cell corresponding to the centre of the box while it is glided, $M_{i}:= \mu_{i}(r_0)/\mu_{i}(r_k)$, $N$ is the number of non zero values of $M_{i}$, and $d$ is the dimension of the physical support $A$.

The curve $f(\alpha)$ against $\alpha$ is the \emph{multifractal spectrum}. The curve $D(q):=\tau(q)/(q-1)$ against $q$ is the curve of \emph{generalized dimensions}. Some particular values of $D_q$ can be related to classical definitions of ``dimension''.

In our context, boxes are defined as Moore neighbourhoods of radius $r$ around each point such that a full box can be included inside the study area. The values $\mu_{i}(r)$ are then the sum of all the lot price whose centroid falls inside each box $i$. The chosen radius $r_0$ was the minimum allowed by the resolution and we averaged the results over several $r_k$ to limit the effect of local inaccuracies in the data. The $q$ range is defined as all values such that the resulting $f(\alpha)$ is non-negative.

As defined above, the $\tau(q)$ values should yield $\tau(1)=0$ and $\tau(0)=d_0$, where $d_0$ is the fractal dimension of the support of the measure $\mu$, so that the spectrum touches the identity line for $q=1$ and so that its maximum is $d_0$, achieved for $q=0$. However, by construction, the multiplier method yields $\tau(1)=-d$ and $\tau(0)=0$, where $d$ is the dimension of the space $A$, hence the addition of $d$ in equation~\ref{taumultp}. Since we find more coherent to obtain $\tau(1)=d_0$ instead of $d$, we have added a rescaling by $d_0/d$ to equations~\ref{taumultp} in order to recover the usual $\tau(1)=0$ and $\tau(0)=d_0$.

\section{Detailed spectrum for each type of city model}\label{app:CityModel}

\begin{figure}
      \centering
      \includegraphics[clip,width=\columnwidth,keepaspectratio]{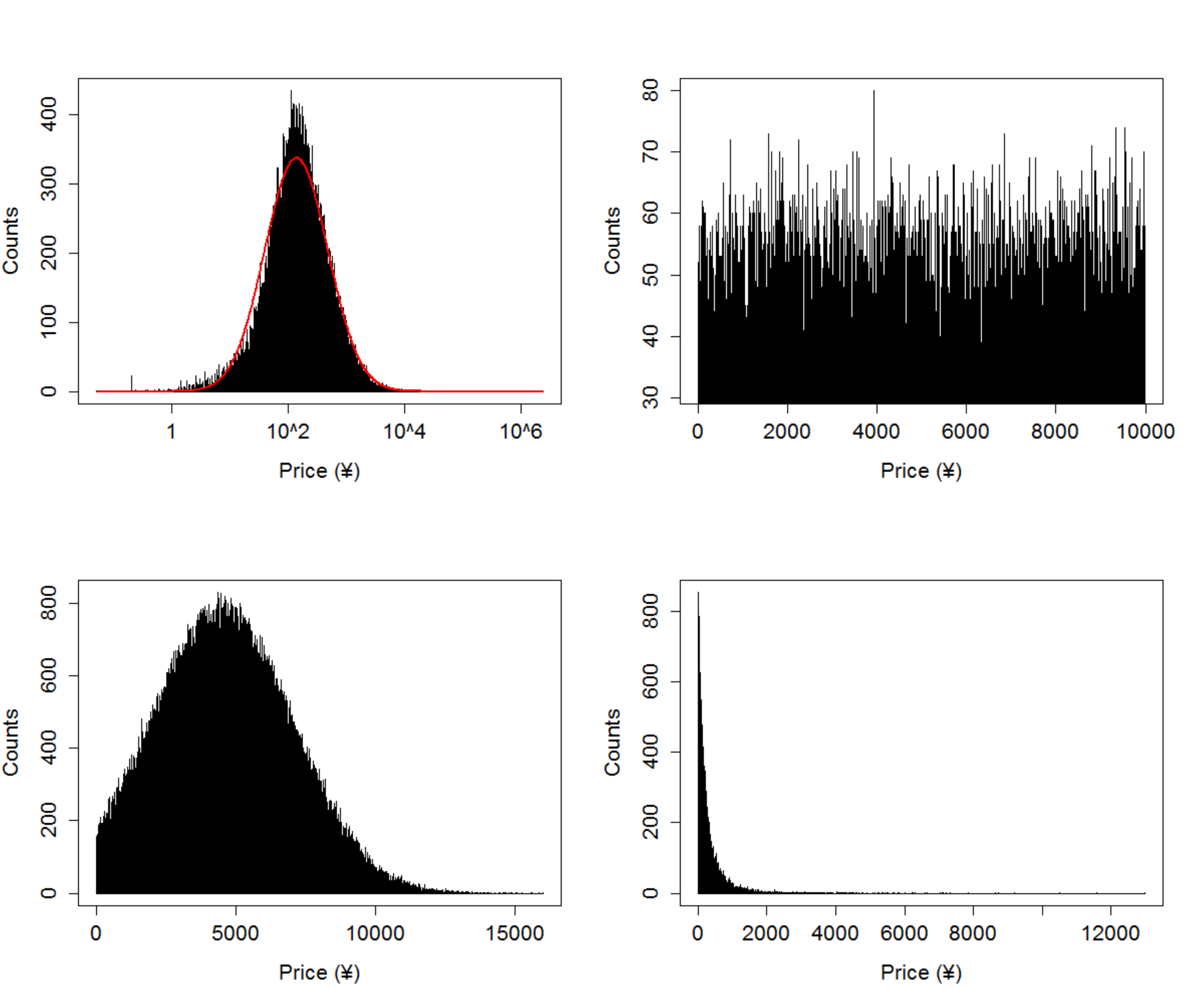}
			\caption{1912 price distributions. Top left: real Kyoto price distribution (note the logarithmic x-axis), top right: uniform distribution, bottom left: truncated normal distribution, bottom right: Pareto distribution.
			\label{abstmodprice}}
\end{figure}

\begin{figure}
      \centering
      \includegraphics[clip,width=\columnwidth,keepaspectratio]{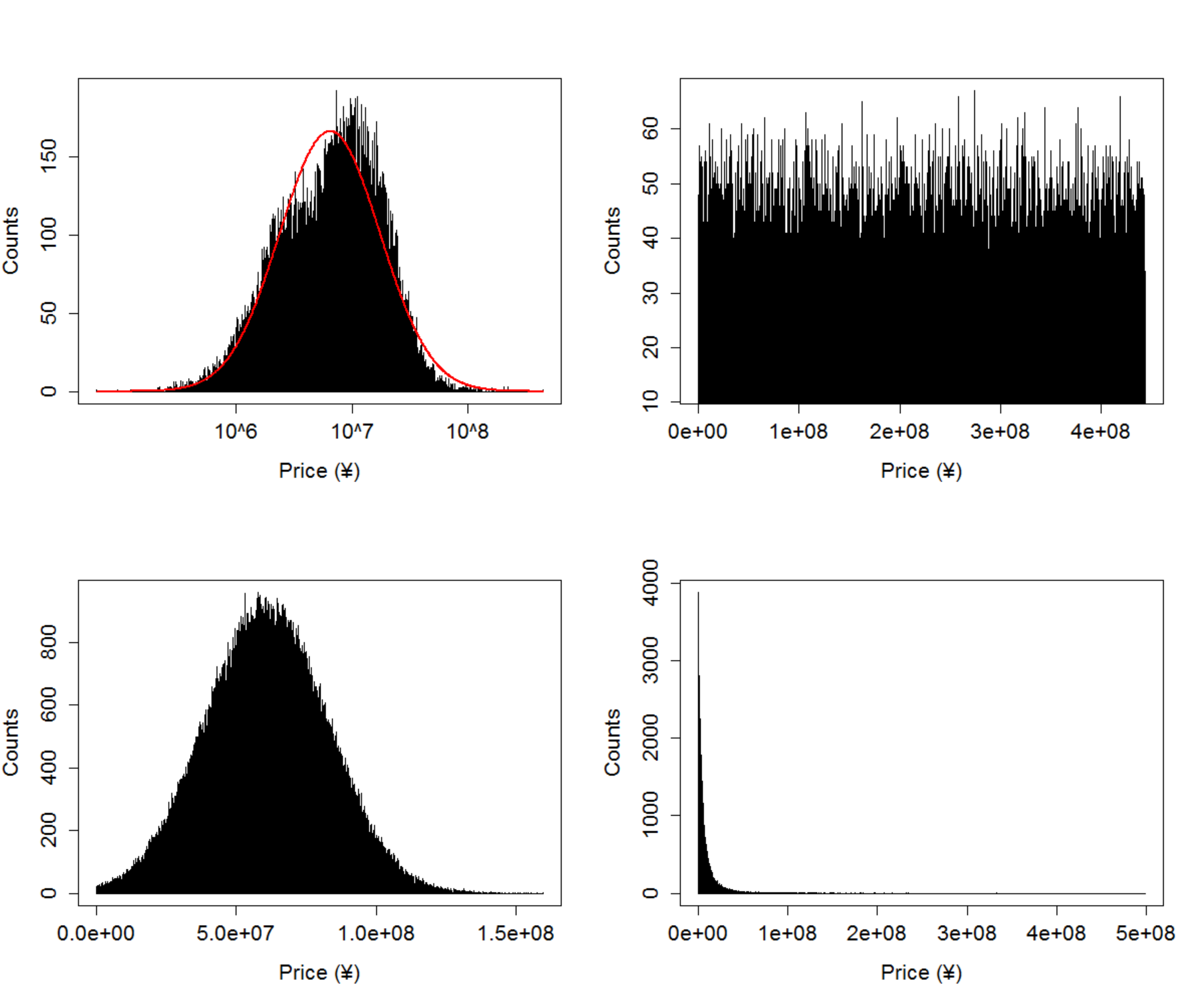}
			\caption{2012 price distributions: real (note the logarithmic x-axis) and corresponding uniform, truncated normal, and Pareto price distributions.
			\label{price2012}}
\end{figure}

The different price distributions used for comparison in the 1912 case are represented in figure~\ref{abstmodprice}, and those for the 2012 case are represented in figure~\ref{price2012}.

\begin{figure}
      \centering
      \includegraphics[clip,width=\columnwidth,keepaspectratio]{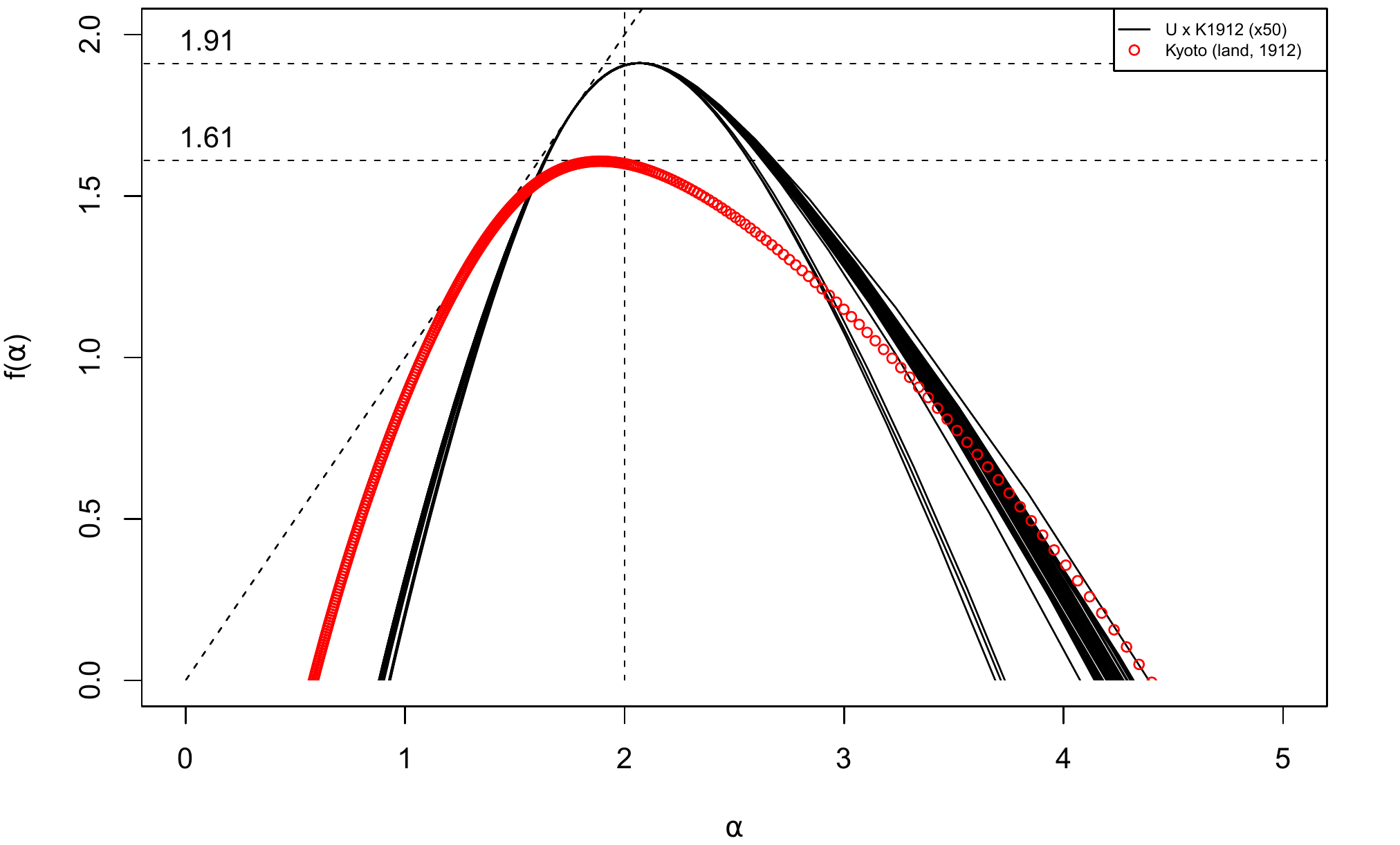}
			\caption{Multifractal spectrum for Kyoto 1912 price distribution drawn uniformly.
			\label{UxR}}
\end{figure}

\begin{figure}
      \centering
      \includegraphics[clip,width=\columnwidth,keepaspectratio]{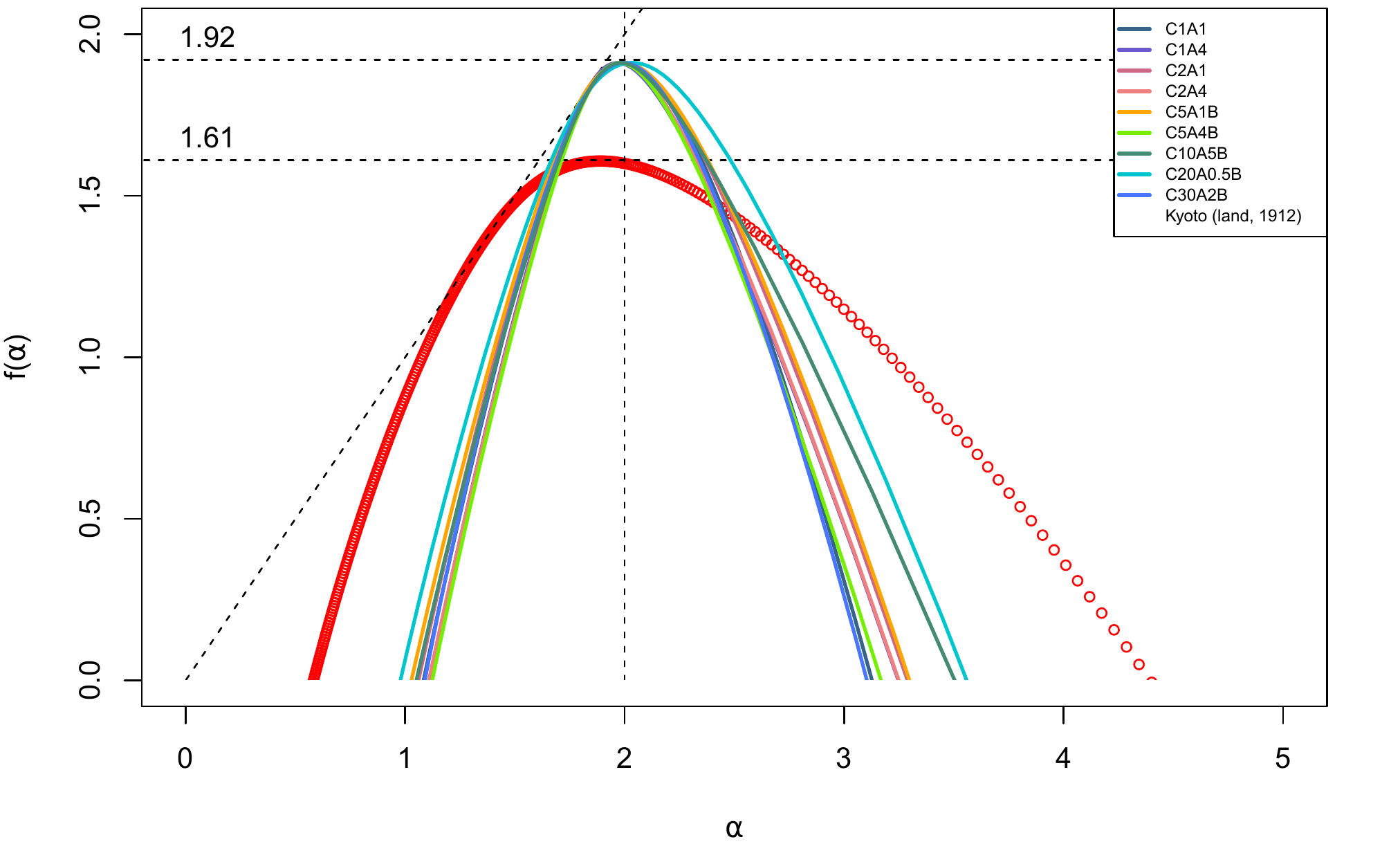}
			\caption{Multifractal spectrum for Kyoto 1912 price distribution mapped on polycentric space distributions. The number after C is the number of centers, the number after A is the global attractivity of centers, and B indicates that centers have different attraction weights.
			\label{PxR}}
\end{figure}

\begin{figure}
      \centering
      \includegraphics[clip,width=\columnwidth,keepaspectratio]{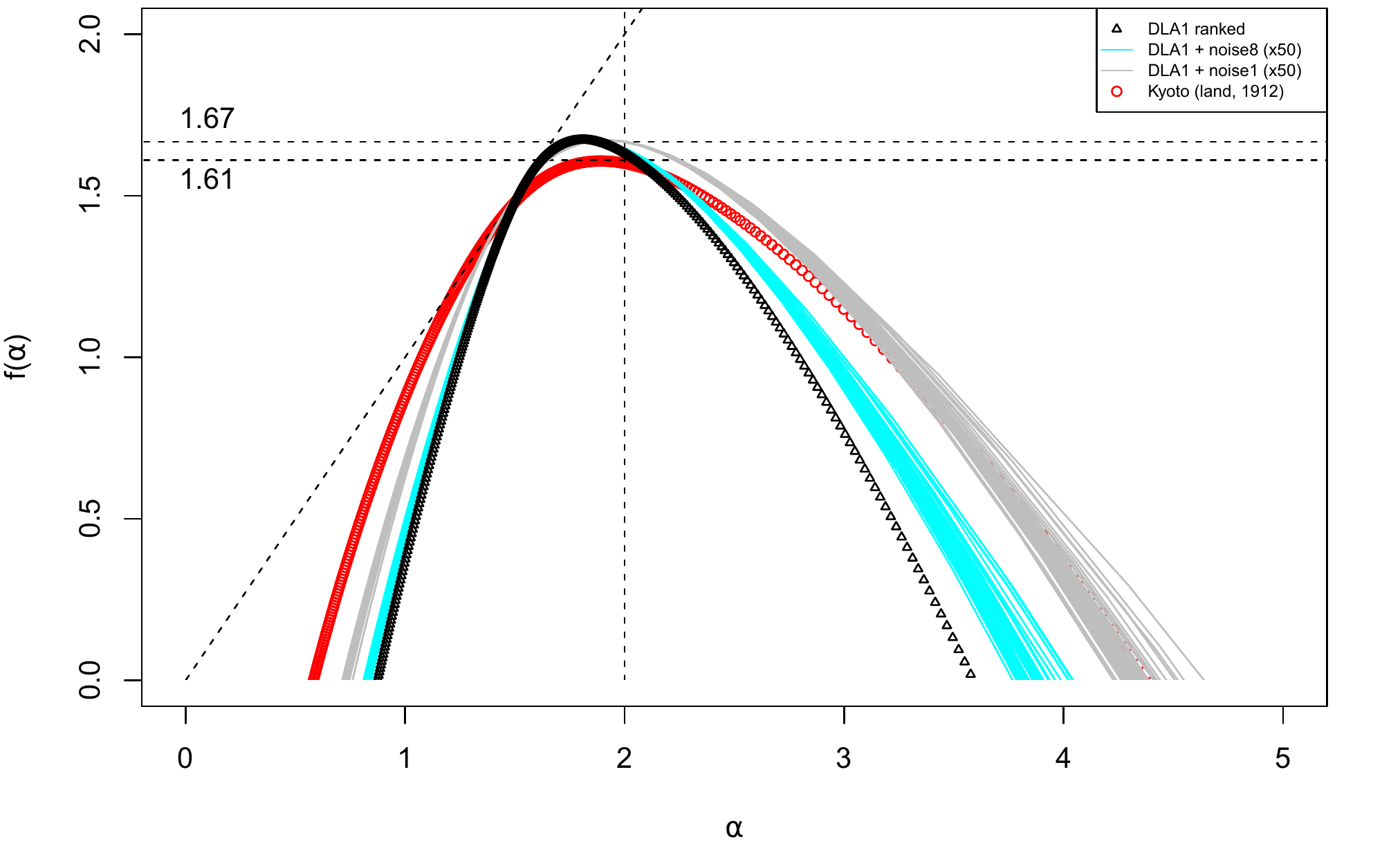}
			\caption{Multifractal spectrum for Kyoto 1912 price distribution mapped on a DLA with 1 center.
			\label{DLA1xR}}
\end{figure}

\begin{figure}
      \centering
      \includegraphics[clip,width=\columnwidth,keepaspectratio]{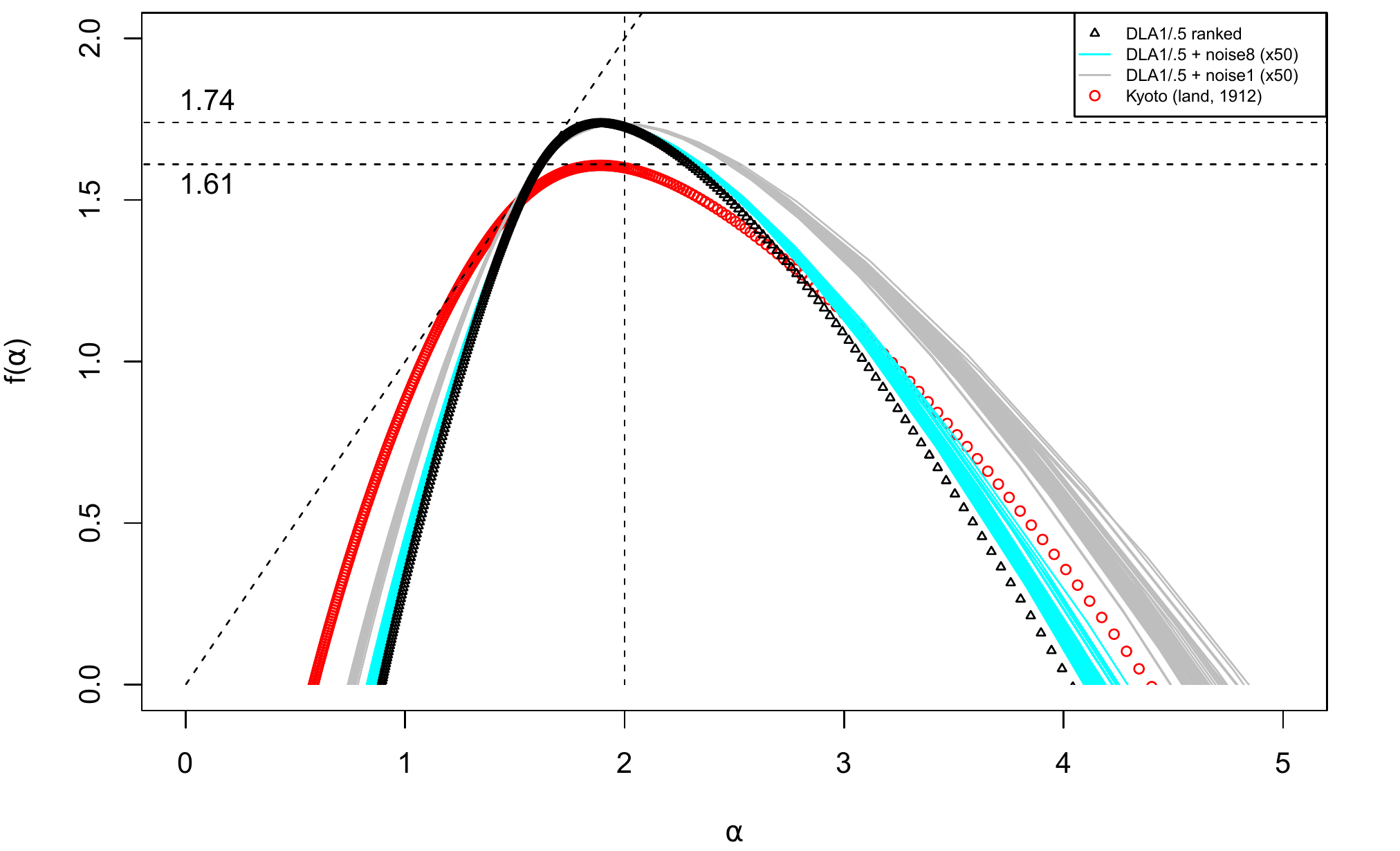}
			\caption{Multifractal spectrum for Kyoto 1912 price distribution mapped on a DLA with 1 center and sticky probability of 0.5.
			\label{DLA15xR}}
\end{figure}

\begin{figure}
      \centering
      \includegraphics[clip,width=\columnwidth,keepaspectratio]{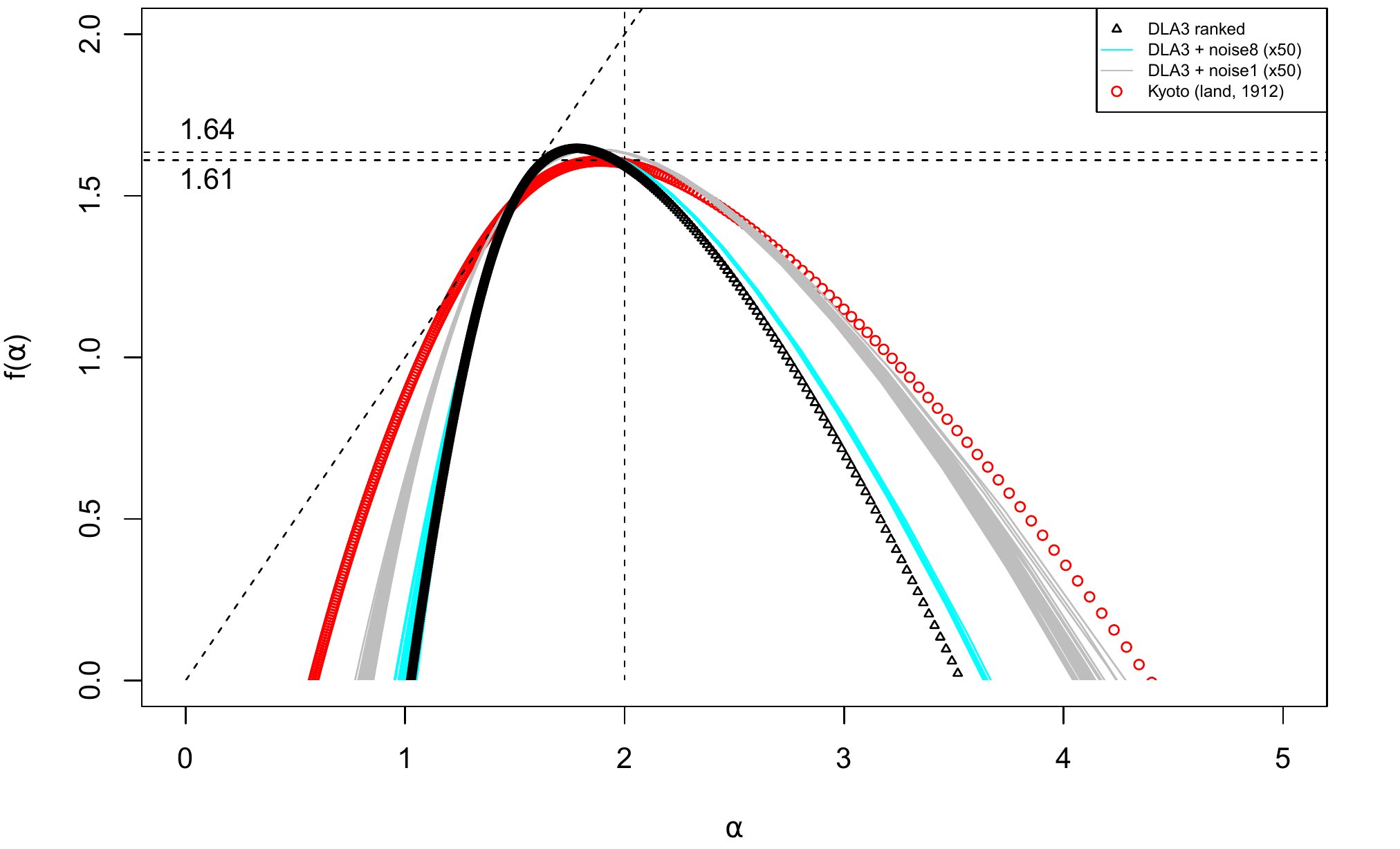}
			\caption{Multifractal spectrum for Kyoto 1912 price distribution mapped on a DLA with 3 centers.
			\label{DLA3xR}}
\end{figure}

\begin{figure}
      \centering
      \includegraphics[clip,width=\columnwidth,keepaspectratio]{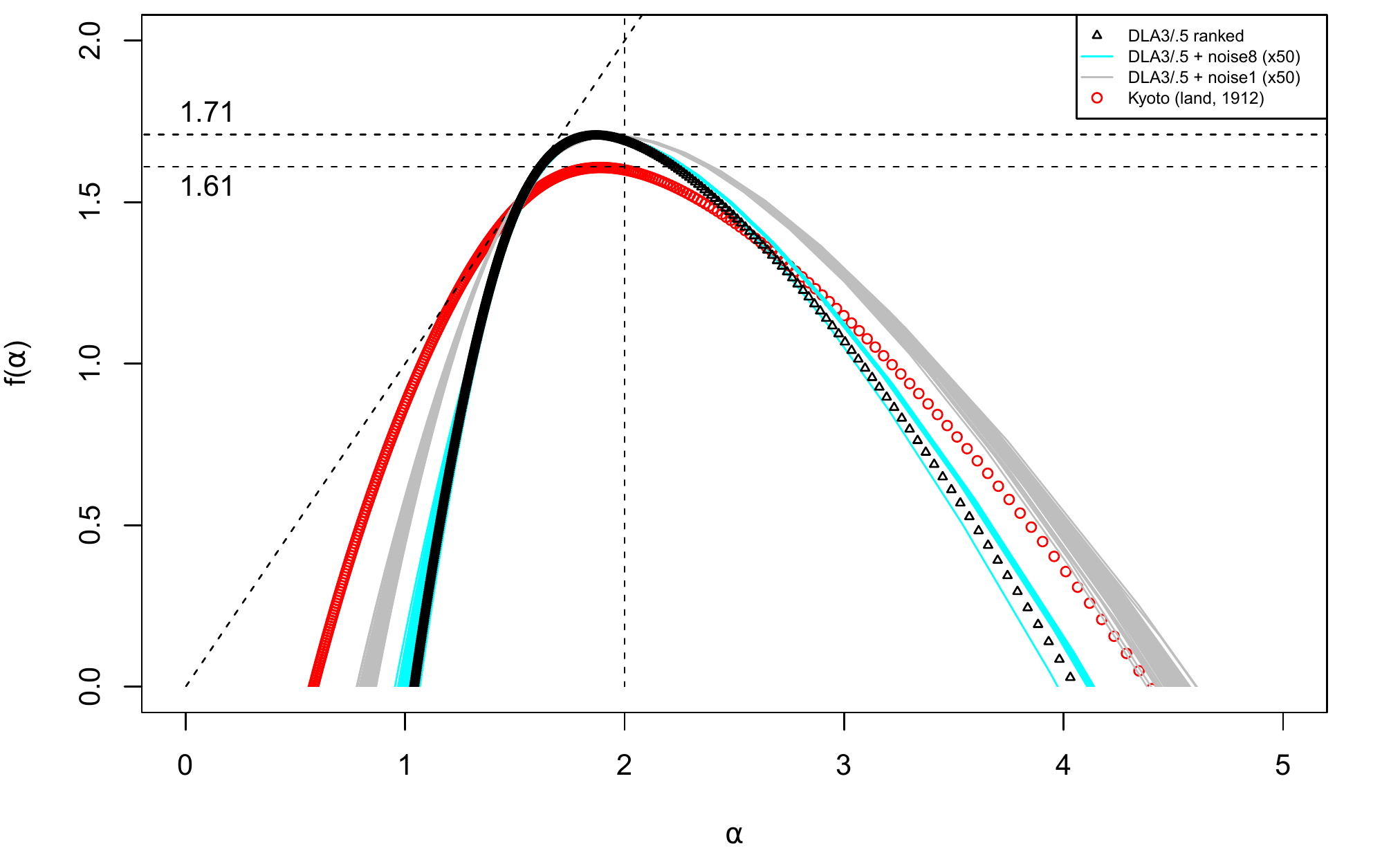}
			\caption{Multifractal spectrum for Kyoto 1912 price distribution mapped on a DLA with 3 centers and sticky probability of 0.5.
			\label{DLA35xR}}
\end{figure}

The uniform distribution (figure~\ref{UxR}) has been drawn 50 times. The DLA distributions (figure~\ref{DLA1xR}, \ref{DLA15xR}, \ref{DLA3xR} \& \ref{DLA35xR}) have first been drawn once in a configuration where the particles representing land lots are liberated into the system according to their rank in the price distribution (most expensive lots being sent first), then two types of noise were added (50 times each) to the price rankings. In the legend, $DLAx(/y)$ refers to a DLA model using $x$ seeds and a sticky coefficient of $y$, Noise$x$ refers to a draw without replacement of the price distribution with a probability distribution of the type $rank^{p}/\sum(rank^{p})$, with $p=x$. For the nine polycentric models (figure~\ref{PxR}), the number after $C$ in the legend is the number of centres, the number after $A$ the global attractivity of the centres, and the letter $B$ indicates that the centres were given different weights.

\begin{figure}
      \centering
      \includegraphics[clip,width=\columnwidth,keepaspectratio]{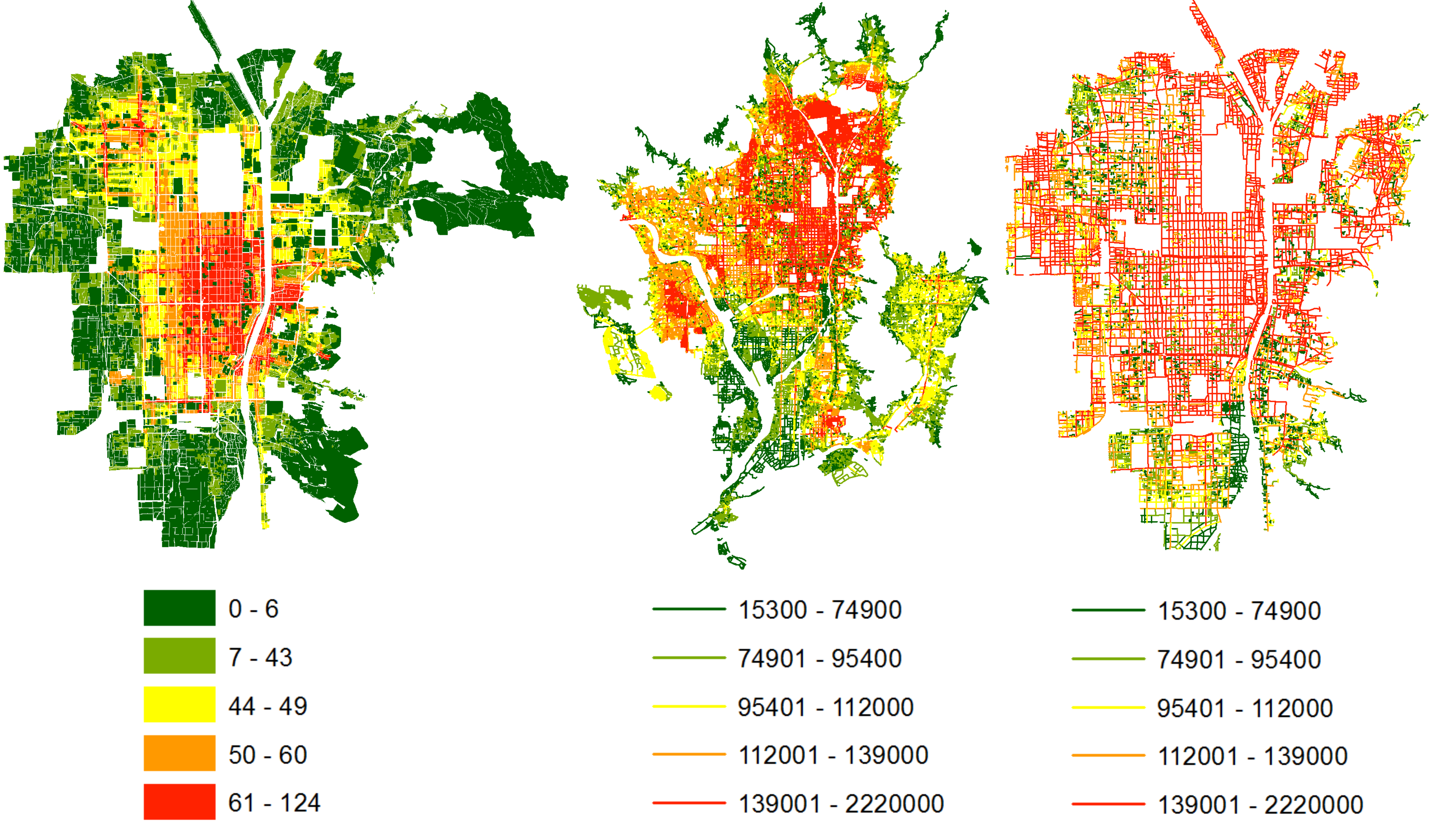}
			\caption{Left: Kyoto 1912 extent; center: Kyoto 2012 extent; right: 2012 Kyoto cropped to the 1912 extent. All prices are expressed in Yen.
			\label{studyareas}}
\end{figure}

The resolution issue mentioned for Kyoto between the 1912 data and 2012 data is illustrated in figure~\ref{studyareas}.

\section{Tables of D0, D1 and D2 for all cases}\label{app:D0D1D2}

Among the generalized dimension defined by the equation $D(q):=\tau(q)/(q-1)$ (see \ref{app:multtheo}), three values relate to well known dimensions: $D(0)$, $D(1)$ and $D(2)$. The first one, $D(0)$, is the fractal dimension of the physical space supporting the measure. The dimension $D(1)$ is the information dimension, it relates to Shannon's entropy and provides a measure of the density evenness in the data. Finally, $D(2)$ is the correlation dimension, which provides a measure of scattering in the data. By definition, the full range of $D(q)$ is the same as the full range of $\alpha$. The $D(0)$ dimension can be computed directly using box-counting, while the $D(1)$ and $D(2)$ values can be deduced from  the direct expressions
\begin{gather}
D(1)=\lim_{r\rightarrow0}\frac{\sum_i\mu_i(r)\log(\mu_i(r))}{\log(r)},\label{D1}\\
D(2)=\lim_{r\rightarrow0}\frac{\log(\sum_i\mu_i(r)^2)}{\log(r)}.\label{D2}
\end{gather}

The $D_0$, $D_1$ and $D_2$ values for Kyoto in 1912, 2012 and 2012 intersected with 1912 boundaries, for the uniform, polycentric and DLA models for 1912 data and shuffled model for 2012 data, as well as for Manhattan assessed tax land value in 2016 and London house price transactions in 2016 can be found in table \ref{tab:D0D1D2}. Those values show the same decrease in multifractality for modern data as the spectra in the main text. Three problematic values with slightly higher $D_1$ than $D_0$ have been obtained for Kyoto in 2012 (1.74 vs 1.70), DLA 1 ranked (1.70 vs 1.67) and DLA 1.5 ranked (1.78 vs 1.74).

\bigskip

\noindent\begin{minipage}{\textwidth}
	\makebox[\textwidth][c]{\small
		\begin{tabular}{@{}|r|c|c|c|}
		\hline
			Model & $D_0$ & $D_1$ & $D_2$ \\ \hline
			Kyoto 1912 & 1.61 & 1.50 & 0.23 \\ \hline
			Kyoto 2012 & 1.70 & 1.74 & 1.70 \\ \hline
			Kyoto 2012 (part.) & 1.72 & 1.70 & 1.58 \\ \hline
			Uniform (1912) & 1.91 & 1.34 & 0.03 \\ \hline
			Polycentric (1912) & 1.91 & 1.34 & 0.03 \\ \hline
			DLA 1 (ranked) & 1.67 & 1.70 & 1.23 \\ \hline
			DLA 1.5 (ranked) & 1.74 & 1.78 & 1.35 \\ \hline
			DLA 1 (noise) & 1.67 & 1.62 & 0.98 \\ \hline
			DLA 1.5 (noise) & 1.74 & 1.54 & 0.70 \\ \hline
			Shuffled (2012) & 1.68 & 1.51 & 1.37 \\ \hline
			Manhattan 2016 & 1.71 & 1.51 & 1.37 \\ \hline
			London 2016 & 1.80 & 1.76 & 1.60 \\ \hline
		\end{tabular}}
	\captionof{table}{$D_0$, $D_1$ and $D_2$ values for the main city models.}
	\label{tab:D0D1D2}
\end{minipage}

\section{Formal definitions of classical inequality indicators}\label{app:CI}

Denote $\{x_i\}_{1\leq i\leq n}$ the set formed by $n$ observations, $\mu$ their average value and $p_i$ the associated probability distribution defined as $p_i=x_i/\sum_j x_j$. Then, the relative dispersion (RD) is defined as
\begin{equation}
\text{RD}=\frac{\sum_i\left|\mu-x_i\right|}{n\mu}.
\end{equation}
The Gini coefficient (G) is defined as
\begin{equation}
\text{G}=\frac{1}{2}\frac{n^2\mu}{\sum_i\sum_j\left|x_i-x_j\right|}.
\end{equation}
The Theil coefficient (T) is defined as
\begin{equation}
\text{T}=\sum_i p_i\log(np_i).
\end{equation}

Now, assume the space is divided into $n$ neighbourhoods, denote $\{n_k\}_{1\leq k\leq n}$ the population count inside each neighbourhood, and $\{\mu_k\}_{1\leq k\leq n}$ the average value inside each neighbourhood. Then, the neighbourhood Sorting Index (NSI) is defined as
\begin{equation}
\text{NSI}=\sqrt{\frac{\frac{\sum_k n_k(\mu_k-\mu)^2}{n}}{{\frac{\sum_i(x_i-\mu)^2}{n}}}}
\end{equation}

For the indices defined by Reardon et al. \cite{RFOM}, it is needed to define first a segregation measure $S$ comparing inter-neighbourhood variation and total variation by
\begin{equation}
S(v)=\sum_{k=1}^N\frac{n_k}{nv}(v-v_k),
\end{equation}
where $v$ is a chosen variation function, and $v_k$ is its value inside neighbourhood $k$. The Ordinal Information Theory Index (OITI) and Ordinal Variation Ratio Index (OVRI) are defined respectively for the following variation functions $v_1$ and $v_2$ 
\begin{gather}
v_1=\frac{1}{K}\sum_{i=1}^K -[c_i\log(c_i)+(1-c_i)\log(1-c_i)];\\
v_2=\frac{1}{K}\sum_{i=1}^K 4c_i(1-c_i),
\end{gather}
where $K$ is the number of ordinal categories considered and $c_i$ is the cumulative proportion of values inside a sample (here, either a singleton or a neighbourhood) of category $i$ or below.

\section*{References}


\begin{thebibliography}{33}
\providecommand{\natexlab}[1]{#1}
\providecommand{\url}[1]{\texttt{#1}}
\providecommand{\urlprefix}{URL }
\expandafter\ifx\csname urlstyle\endcsname\relax
  \providecommand{\doi}[1]{doi:\discretionary{}{}{}#1}\else
  \providecommand{\doi}[1]{doi:\discretionary{}{}{}\begingroup
  \urlstyle{rm}\url{#1}\endgroup}\fi
\providecommand{\bibinfo}[2]{#2}

\bibitem[{Reardon et~al.(????)Reardon, Firebaugh, O'Sullivan, and
  Mathews}]{RFOM}
\bibinfo{author}{S.~Reardon}, \bibinfo{author}{G.~Firebaugh},
  \bibinfo{author}{D.~O'Sullivan}, \bibinfo{author}{S.~Mathews},
  \bibinfo{title}{A New Approach to Measuring Socio-Spatial Economic
  Segregation}, \bibinfo{journal}{Prepared for the 29th General Conference of
  The International Association for Research in Income and Wealth}
  \bibinfo{volume}{(2006)}.

\bibitem[{Frankhauser(1994)}]{Fh}
\bibinfo{author}{P.~Frankhauser}, \bibinfo{title}{La fractalite des structures
  urbaines}, \bibinfo{publisher}{Anthropos}, \bibinfo{address}{Paris},
  \bibinfo{year}{1994}.

\bibitem[{Batty and Longley(1994)}]{BL}
\bibinfo{editor}{M.~Batty}, \bibinfo{editor}{P.~Longley} (Eds.),
  \bibinfo{title}{Fractal cities: a geometry of form and function},
  \bibinfo{publisher}{CA: Academic Press}, \bibinfo{address}{San Diego},
  \bibinfo{year}{1994}.

\bibitem[{Batty(2008)}]{B}
\bibinfo{author}{M.~Batty}, \bibinfo{title}{The size, scale and shape of
  cities}, \bibinfo{journal}{Science} \bibinfo{volume}{319}
  (\bibinfo{year}{2008}) \bibinfo{pages}{769--771},
  \urlprefix\url{http://dx.doi.org/10.1126/science.1151419}.

\bibitem[{Hu et~al.(2012)Hu, Cheng, Wang, and Xie}]{HCWX}
\bibinfo{author}{S.~Hu}, \bibinfo{author}{Q.~Cheng}, \bibinfo{author}{L.~Wang},
  \bibinfo{author}{S.~Xie}, \bibinfo{title}{Multifractal characterization of
  urban residential land price in space and time}, \bibinfo{journal}{Appl.
  Geogr.} \bibinfo{volume}{34} (\bibinfo{year}{2012})
  \bibinfo{pages}{161--170},
  \urlprefix\url{http://dx.doi.org/10.1016/j.apgeog.2011.10.016}.

\bibitem[{Hu et~al.(2013)Hu, Cheng, Wang, and Xu}]{HCWX2}
\bibinfo{author}{S.~Hu}, \bibinfo{author}{Q.~Cheng}, \bibinfo{author}{L.~Wang},
  \bibinfo{author}{D.~Xu}, \bibinfo{title}{Modeling land price distribution
  using multifractal {IDW} interpolation and fractal filtering method},
  \bibinfo{journal}{Landscape Urban Plan.} \bibinfo{volume}{110}
  (\bibinfo{year}{2013}) \bibinfo{pages}{25--35},
  \urlprefix\url{http://dx.doi.org/10.1016/j.landurbplan.2012.09.008}.

\bibitem[{YNI(2007)}]{YNIBS}
\bibinfo{title}{Meiji-Taisho Era in Virtual Kyoto (1868-1926)}, in:
  \bibinfo{editor}{K.~Yano}, \bibinfo{editor}{T.~Nakaya},
  \bibinfo{editor}{Y.~Isoda} (Eds.), \bibinfo{booktitle}{Virtual Kyoto:
  Exploring the past, present and future of Kyoto}, chap.~\bibinfo{chapter}{3},
  \bibinfo{publisher}{Nakanishiya Publishing Company}, \bibinfo{pages}{48--69},
  \bibinfo{year}{2007}.

\bibitem[{Yano et~al.(2008)Yano, Nakaya, Isoda, Takase, Kawasumi, Matsuoka,
  Seto, Kawahara, Tsukamoto, Inoue, and Kirimura}]{YNITKMSKTIK}
\bibinfo{author}{K.~Yano}, \bibinfo{author}{T.~Nakaya},
  \bibinfo{author}{Y.~Isoda}, \bibinfo{author}{Y.~Takase},
  \bibinfo{author}{T.~Kawasumi}, \bibinfo{author}{K.~Matsuoka},
  \bibinfo{author}{T.~Seto}, \bibinfo{author}{D.~Kawahara},
  \bibinfo{author}{A.~Tsukamoto}, \bibinfo{author}{M.~Inoue},
  \bibinfo{author}{T.~Kirimura}, \bibinfo{title}{Virtual Kyoto: 4D-GIS
  comprising spatial and temporal dimensions}, \bibinfo{journal}{J. Geogr.}
  \bibinfo{volume}{117} (\bibinfo{year}{2008}) \bibinfo{pages}{464--478},
  \urlprefix\url{http://dx.doi.org/10.5026/jgeography.117.464}.

\bibitem[{Yano et~al.(2009)Yano, Nakaya, Isoda, and Kawasumi}]{YNIK}
\bibinfo{author}{K.~Yano}, \bibinfo{author}{T.~Nakaya},
  \bibinfo{author}{Y.~Isoda}, \bibinfo{author}{T.~Kawasumi},
  \bibinfo{title}{Virtual Kyoto as {4D GIS}}, in: \bibinfo{editor}{H.~Lin},
  \bibinfo{editor}{M.~Batty} (Eds.), \bibinfo{booktitle}{Virtual Geographic
  Environments}, \bibinfo{publisher}{Science Press}, \bibinfo{pages}{71--88},
  \bibinfo{year}{2009}.

\bibitem[{Yano et~al.(2011)Yano, Nakaya, Kawasumi, and Tanaka}]{YNKT}
\bibinfo{editor}{K.~Yano}, \bibinfo{editor}{T.~Nakaya},
  \bibinfo{editor}{T.~Kawasumi}, \bibinfo{editor}{S.~Tanaka} (Eds.),
  \bibinfo{title}{Historical {GIS} of Kyoto}, \bibinfo{publisher}{Nakanishiya
  Publishing Company}, \bibinfo{year}{2011}.

\bibitem[{Murcio et~al.(2015)Murcio, Masucci, Arcaute, and Batty}]{MMAB}
\bibinfo{author}{R.~Murcio}, \bibinfo{author}{A.~P. Masucci},
  \bibinfo{author}{E.~Arcaute}, \bibinfo{author}{M.~Batty},
  \bibinfo{title}{Multifractal to monofractal evolution of the London's street
  network}, \bibinfo{journal}{Phys. Rev. E} \bibinfo{volume}{92}
  (\bibinfo{year}{2015}) \bibinfo{pages}{062130},
  \urlprefix\url{http://dx.doi.org/10.1103/PhysRevE.92.062130}.

\bibitem[{Ariza-Villaverde et~al.(2013)Ariza-Villaverde, Jimenez-Hornero, and
  Rave}]{AVHG}
\bibinfo{author}{A.~B. Ariza-Villaverde}, \bibinfo{author}{F.~J.
  Jimenez-Hornero}, \bibinfo{author}{E.~G.~D. Rave},
  \bibinfo{title}{Multifractal analysis of axial maps applied to the study of
  urban morphology}, \bibinfo{journal}{Comput. Environ. Urban}
  \bibinfo{volume}{38} (\bibinfo{year}{2013}) \bibinfo{pages}{1--10},
  \urlprefix\url{http://dx.doi.org/10.1016/j.compenvurbsys.2012.11.001}.

\bibitem[{Frisch and Parisi(????)}]{FP}
\bibinfo{author}{U.~Frisch}, \bibinfo{author}{G.~Parisi},
  \bibinfo{title}{Turbulence and Predictability of Geophysical Flows and
  Climate Dynamics}, \bibinfo{journal}{Proc. Varenna Summer School LXXXVIII}
  \bibinfo{volume}{(1983)}.

\bibitem[{Evertsz and Mandelbrot(1992)}]{EM}
\bibinfo{author}{C.~J. Evertsz}, \bibinfo{author}{B.~B. Mandelbrot},
  \bibinfo{title}{Multifractal Measures}, in: \bibinfo{editor}{H.~Peitgen},
  \bibinfo{editor}{H.~J{\"u}rgens}, \bibinfo{editor}{D.~Saupe} (Eds.),
  \bibinfo{booktitle}{Chaos and Fractals}, \bibinfo{publisher}{Springer},
  \bibinfo{address}{New York}, \bibinfo{pages}{849--881}, \bibinfo{year}{1992}.

\bibitem[{Falconer(2003)}]{F}
\bibinfo{author}{K.~Falconer}, \bibinfo{title}{Fractal Geometry,
  2\textsuperscript{nd} Ed.}, \bibinfo{publisher}{Wiley},
  \bibinfo{address}{Chichester, West Sussex, England}, \bibinfo{year}{2003}.

\bibitem[{Salat et~al.(2017)Salat, Murcio, and Arcaute}]{SMA}
\bibinfo{author}{H.~Salat}, \bibinfo{author}{R.~Murcio},
  \bibinfo{author}{E.~Arcaute}, \bibinfo{title}{Multifractal Methodology},
  \bibinfo{journal}{Physica A} \bibinfo{volume}{47} (\bibinfo{year}{2017})
  \bibinfo{pages}{467--487},
  \urlprefix\url{http://dx.doi.org/10.1016/j.physa.2017.01.041}.

\bibitem[{Halsey et~al.(1986)Halsey, Jensen, Kadanoff, Procaccia, and
  Shraiman}]{HJKPS}
\bibinfo{author}{T.~C. Halsey}, \bibinfo{author}{M.~H. Jensen},
  \bibinfo{author}{L.~P. Kadanoff}, \bibinfo{author}{I.~Procaccia},
  \bibinfo{author}{B.~I. Shraiman}, \bibinfo{title}{Fractal measures and their
  singularities: The characterization of strange sets}, \bibinfo{journal}{Phys.
  Rev. A} \bibinfo{volume}{33} (\bibinfo{year}{1986})
  \bibinfo{pages}{1141--1151},
  \urlprefix\url{http://dx.doi.org/10.1103/PhysRevA.33.1141}.

\bibitem[{Atmanspacher et~al.(1989)Atmanspacher, Scheingraber, and
  Wiedenmann}]{ASW}
\bibinfo{author}{H.~Atmanspacher}, \bibinfo{author}{H.~Scheingraber},
  \bibinfo{author}{G.~Wiedenmann}, \bibinfo{title}{Determination of $f(\alpha)$
  for a limited random point set}, \bibinfo{journal}{Phys. Rev. A}
  \bibinfo{volume}{40} (\bibinfo{year}{1989}) \bibinfo{pages}{3954--3963},
  \urlprefix\url{http://dx.doi.org/10.1103/PhysRevA.40.3954}.

\bibitem[{Chhabra and Sreenivasan(1991)}]{CS}
\bibinfo{author}{A.~B. Chhabra}, \bibinfo{author}{K.~Sreenivasan},
  \bibinfo{title}{Negative dimensions: Theory, computation, and experiment},
  \bibinfo{journal}{Phys. Rev. A} \bibinfo{volume}{43} (\bibinfo{year}{1991})
  \bibinfo{pages}{1114--1117},
  \urlprefix\url{http://dx.doi.org/10.1103/PhysRevA.43.1114}.

\bibitem[{Cheng(1999)}]{C}
\bibinfo{author}{Q.~Cheng}, \bibinfo{title}{The gliding box method for
  multifractal modeling}, \bibinfo{journal}{Comput. Geosci.}
  \bibinfo{volume}{25} (\bibinfo{year}{1999}) \bibinfo{pages}{1073--1079},
  \urlprefix\url{http://dx.doi.org/10.1016/S0098-3004(99)00068-0}.

\bibitem[{Odland(1978)}]{O2}
\bibinfo{author}{J.~Odland}, \bibinfo{title}{The Conditions for Multi-Center
  Cities}, \bibinfo{journal}{Econ. Geogr.} \bibinfo{volume}{54}
  (\bibinfo{year}{1978}) \bibinfo{pages}{234--244},
  \urlprefix\url{http://dx.doi.org/10.2307/142837}.

\bibitem[{McMillen and Smith(2003)}]{MS}
\bibinfo{author}{D.~P. McMillen}, \bibinfo{author}{S.~C. Smith},
  \bibinfo{title}{The number of subcenters in marge urban areas},
  \bibinfo{journal}{J. Urban Econ.} \bibinfo{volume}{53} (\bibinfo{year}{2003})
  \bibinfo{pages}{321--338},
  \urlprefix\url{http://dx.doi.org/10.1016/S0094-1190(03)00026-3}.

\bibitem[{Louf and Barthelemy(2013)}]{LB}
\bibinfo{author}{R.~Louf}, \bibinfo{author}{M.~Barthelemy},
  \bibinfo{title}{Modeling the Polycentric Transition of Cities},
  \bibinfo{journal}{Phys. Rev. Lett.} \bibinfo{volume}{111}
  (\bibinfo{year}{2013}) \bibinfo{pages}{198702},
  \urlprefix\url{http://dx.doi.org/10.1103/PhysRevLett.111.198702}.

\bibitem[{Witten and Sander(1981)}]{WS}
\bibinfo{author}{T.~A. Witten}, \bibinfo{author}{L.~M. Sander},
  \bibinfo{title}{Diffusion-Limited Aggregation, a Kinetic Critical
  Phenomenon}, \bibinfo{journal}{Phys. Rev. Lett.} \bibinfo{volume}{47}
  (\bibinfo{year}{1981}) \bibinfo{pages}{203--218},
  \urlprefix\url{http://dx.doi.org/10.1080/001075100409698}.

\bibitem[{Vicsek et~al.(1990)Vicsek, Family, and Meakin}]{VFM}
\bibinfo{author}{T.~Vicsek}, \bibinfo{author}{F.~Family},
  \bibinfo{author}{P.~Meakin}, \bibinfo{title}{Multifractal Geometry of
  Diffusion-Limited Aggregates}, \bibinfo{journal}{Europhys. Lett.}
  \bibinfo{volume}{12} (\bibinfo{year}{1990}) \bibinfo{pages}{217--222},
  \urlprefix\url{http://dx.doi.org/10.1209/0295-5075/12/3/005}.

\bibitem[{Murcio and Rodriguez-Romo(2009)}]{MRR}
\bibinfo{author}{R.~Murcio}, \bibinfo{author}{S.~Rodriguez-Romo},
  \bibinfo{title}{Colored diffusion-limited aggregation for urban migration},
  \bibinfo{journal}{Physica A} \bibinfo{volume}{388} (\bibinfo{year}{2009})
  \bibinfo{pages}{2689--2698},
  \urlprefix\url{http://dx.doi.org/10.1016/j.physa.2009.03.021}.

\bibitem[{Rodriguez-Romo and Murcio(2014)}]{RRM}
\bibinfo{author}{S.~Rodriguez-Romo}, \bibinfo{author}{R.~Murcio},
  \bibinfo{title}{An assessment of similarity measures for aggregates grown
  from multiple seeds}, \bibinfo{journal}{Chaos Soliton Fract.}
  \bibinfo{volume}{66} (\bibinfo{year}{2014}) \bibinfo{pages}{31--40},
  \urlprefix\url{http://dx.doi.org/10.1016/j.physa.2009.03.021}.

\bibitem[{Apparicio et~al.(2013)Apparicio, Martori, Pearson, Fournier, and
  Apparicio}]{AMPFA}
\bibinfo{author}{P.~Apparicio}, \bibinfo{author}{J.~C. Martori},
  \bibinfo{author}{A.~L. Pearson}, \bibinfo{author}{E.~Fournier},
  \bibinfo{author}{D.~Apparicio}, \bibinfo{title}{An Open-Source Software for
  Calculating Indices of Urban Residential Segregation}, \bibinfo{journal}{Sac.
  Sci. Comput. Rev.} \bibinfo{volume}{32} (\bibinfo{year}{2013})
  \bibinfo{pages}{117--128},
  \urlprefix\url{http://dx.doi.org/10.1177/0894439313504539}.

\bibitem[{Jargowsky(1996)}]{J}
\bibinfo{author}{P.~A. Jargowsky}, \bibinfo{title}{Take the Money and Run:
  Economic Segregation in U.S. Metropolitan Areas}, \bibinfo{journal}{Am.
  Sociol. Rev.} \bibinfo{volume}{61} (\bibinfo{year}{1996})
  \bibinfo{pages}{984--998}, \urlprefix\url{http://dx.doi.org/10.2307/2096304}.

\bibitem[{Chakravorty(1996)}]{C2}
\bibinfo{author}{S.~Chakravorty}, \bibinfo{title}{A measurement of Spatial
  Disparity: The Case of Income Inequality}, \bibinfo{journal}{Urban Stud.}
  \bibinfo{volume}{33} (\bibinfo{year}{1996}) \bibinfo{pages}{1671--1686},
  \urlprefix\url{http://dx.doi.org/10.1080/0042098966556}.

\bibitem[{Sen(1973)}]{S}
\bibinfo{author}{A.~Sen}, \bibinfo{title}{On Economic Inequality},
  \bibinfo{publisher}{Clarendon Press}, \bibinfo{address}{Oxford},
  \bibinfo{note}{expanded edition with an annex ``On Economic Inequality after
  a Quarter Century'' [jointly with James Foster], Oxford: Clarendon Press.},
  \bibinfo{year}{1973}.

\bibitem[{Fujita and Hill(1997)}]{FCH}
\bibinfo{author}{K.~Fujita}, \bibinfo{author}{R.~C. Hill},
  \bibinfo{title}{Together and equal: place stratification in Osaka}, in:
  \bibinfo{editor}{P.~P. Karan}, \bibinfo{editor}{K.~Stapleton} (Eds.),
  \bibinfo{booktitle}{The Japanese city}, \bibinfo{publisher}{Springer},
  \bibinfo{address}{Kentucky UP, Lexington}, \bibinfo{pages}{106--133},
  \bibinfo{year}{1997}.

\bibitem[{Fielding(2004)}]{Fd}
\bibinfo{author}{A.~J. Fielding}, \bibinfo{title}{Class and space: social
  segregation in Japanese cities}, \bibinfo{journal}{Trans. Inst. Br. Geogr.}
  \bibinfo{volume}{29} (\bibinfo{year}{2004}) \bibinfo{pages}{64--84},
  \urlprefix\url{http://dx.doi.org/10.1111/j.0020-2754.2004.00114.x}.

\end{thebibliography}

\end{document}